\documentclass[lettersize,journal]{IEEEtran}
\usepackage{amsmath,amsfonts}
\usepackage[indLines=false, italicComments=false, commentColor=black]{algpseudocodex}
\usepackage{algorithm}
\usepackage{array}
\usepackage[caption=false,font=normalsize,labelfont=sf,textfont=sf]{subfig}
\usepackage{textcomp}
\usepackage{stfloats}
\usepackage{url}
\usepackage{verbatim}
\usepackage{graphicx}
\usepackage{cite}
\usepackage{booktabs}
\usepackage{amsmath} 
\usepackage{amsfonts} 
\usepackage[normalem]{ulem} 

\usepackage{multirow}
\usepackage{eso-pic}
\usepackage{mwe}

\algnewcommand\algorithmicforeach{\textbf{for each}}
\algdef{S}[FOR]{ForEach}[1]{\algorithmicforeach\ #1\ \algorithmicdo}

\counterwithout{algorithm}{section}

\newcommand{\bigparallel}{\mathop{\parallel}\displaylimits}

\begin{document}

\title{RouteNet-Gauss: Hardware-Enhanced Network Modeling with Machine Learning}

\author{Carlos Güemes-Palau, Miquel Ferriol-Galmés, Jordi Paillisse-Vilanova, Albert López-Brescó, Pere Barlet-Ros, Albert Cabellos-Aparicio
\thanks{All authors are with the Barcelona Neural Networking Center, Universitat Politècnica de Catalunya, Barcelona, Catalonia, Spain.}
\thanks{Manuscript received January 15, 2025.}}

\markboth{IEEE/ACM Transactions on Networking Special Issue on AI and Networking, January 2025}%
{Carlos Güemes Palau, Miquel Ferriol Galmés, \MakeLowercase{\textit{et al.}}: RouteNet-Gauss: Hardware-Enhanced Network Modeling with Machine Learning}

\IEEEpubid{0000--0000/00\$00.00~\copyright~2025 IEEE}

\maketitle

\begin{abstract}
Network simulation is pivotal in network modeling, assisting with tasks ranging from capacity planning to performance estimation. Traditional approaches such as Discrete Event Simulation (DES) face limitations in terms of computational cost and accuracy. This paper introduces RouteNet-Gauss, a novel integration of a testbed network with a Machine Learning (ML) model to address these challenges. By using the testbed as a hardware accelerator, RouteNet-Gauss generates training datasets rapidly and simulates network scenarios with high fidelity to real-world conditions. Experimental results show that RouteNet-Gauss significantly reduces prediction errors by up to 95\% and achieves a 488x speedup in inference time compared to state-of-the-art DES-based methods. RouteNet-Gauss's modular architecture is dynamically constructed based on the specific characteristics of the network scenario, such as topology and routing. 
This enables it to understand and generalize to different network configurations beyond those seen during training, including networks up to 10x larger. Additionally, it supports Temporal Aggregated Performance Estimation (TAPE), providing configurable temporal granularity and maintaining high accuracy in flow performance metrics. This approach shows promise in improving both simulation efficiency and accuracy, offering a valuable tool for network operators.
\end{abstract}

\begin{IEEEkeywords}
Network Modeling, Neural Networks, Graph Neural Networks, Spatio-Temporal Graph Neural Networks.
\end{IEEEkeywords}

\AddToShipoutPictureBG{
  \AtPageUpperLeft{%
    \raisebox{-1.7\baselineskip}{\makebox[\paperwidth]{\begin{minipage}{19cm}
    \footnotesize \textbf{NOTE:} This article has been accepted for publication in IEEE Transactions on Networking. This is the author's version which has not been fully edited and content may change prior to final publication. Citation information: DOI 10.1109/TON.2026.3668972
    \end{minipage}}}%
  }
}
\AddToShipoutPictureBG{
  \AtPageLowerLeft{%
    \raisebox{2\baselineskip}{\makebox[\paperwidth]{\begin{minipage}{19cm}
    \footnotesize © 2026 IEEE. All rights reserved, including rights for text and data mining and training of artificial intelligence and similar technologies. Personal use is permitted, but republication/redistribution requires IEEE permission. See https://www.ieee.org/publications/rights/index.html for more information.
    \end{minipage}}}%
  }
}

\section{Introduction}
\label{sec:intro}
In recent years, there has been significant progress in the development of accurate network models, which have become essential tools for network operators. These models play a pivotal role in facilitating various tasks, including capacity planning \cite{network-planning-21}, topology design \cite{topology-discovery18}, and traffic engineering \cite{auto-dc-cong-control-18}. By providing a controlled and safe environment for testing network configurations, these models eliminate the need to modify production networks directly. This capability enables researchers and operators to explore scenarios that would otherwise be too risky or impractical to test in operational environments.


A widely adopted approach to building these network models involves the use of Discrete Event Simulation (DES) methodologies, such as ns-3~\cite{Riley2010} and OMNeT++~\cite{Varga2019}. DES-based models are highly valued for their ability to provide packet-level visibility, enabling detailed tracking of individual packets as they traverse a network. This level of granularity has made DES methodologies a standard in a wide range of applications, including protocol debugging, performance estimation, topology design, and SLA assurance. However, these models also face notable challenges (explored more thoroughly in Section \ref{sec:background}):
\begin{enumerate} 
    \item \textit{Issue \#1 - Computational Complexity}: DES-based simulators are computationally intensive, as they simulate each individual event in a network (e.g., packet transmission), which can become prohibitive in scenarios involving millions or billions of packets per second \cite{10.1145/3452296.3472926}. 
    \item \textit{Issue \#2 - Potential Inaccuracies}: DES-based simulators often rely on idealized assumptions and scenarios, which may not fully capture the complexities of real-world networks~\cite{216073, 10.1145/3508026}. Furthermore, the lack of access to proprietary details, such as the precise configurations of commercial hardware devices (e.g., queue sizes), potentially limits their accuracy. 
\end{enumerate}


Significant research efforts have focused on reducing the computational cost of DES-based simulations. For example, novel DES designs like DONS~\cite{10.1145/3603269.3604844} have introduced parallelization techniques, achieving remarkable speedups - up to 65 times faster than the standard OMNeT++. Similarly, hybrid approaches such as DeepQueueNet (DQN)~\cite{10.1145/3544216.3544248} leverage Machine Learning (ML) models to accelerate specific components of the simulation pipeline, offering comparable efficiency improvements. These advances represent significant steps toward improving simulation performance.
However, these methods do not address \textit{Issue \#2}, and are therefore constrained by the accuracy ceiling of DES itself. This includes ML models trained on simulated data~\cite{10.1145/3603269.3604844, 10.1145/3452296.3472926, 10.1145/3544216.3544248, ferriolgalmés2022routenetfermi}.

In this paper, we propose a hybrid approach that combines a network testbed with Machine Learning (ML) to address both \textit{Issue \#1} and \textit{Issue \#2}. By leveraging real data from a network testbed, we aim to enhance the accuracy of network modeling and bridge the gap inherent in DES-based solutions (\textit{Issue \#2}). Additionally, the testbed's capability to execute network scenarios \textit{at line rate} enables the model to capture the performance characteristics of high-speed traffic while maintaining efficient inference times (\textit{Issue \#1}). \IEEEpubidadjcol

This approach, however, involves a trade-off. To ensure computational efficiency when modeling high-speed traffic, our method operates at \textit{flow-level granularity} rather than \textit{packet-level granularity}. While many existing solutions rely on packet-level analysis, flow-level granularity can effectively meet the requirements of various network operation tasks, such as capacity planning, QoS assurance, topology design, traffic engineering, and performance estimation~\cite{10.1145/3452296.3472926, ferriol2022routeneterlang, ferriolgalmés2022routenetfermi}. This simplification is particularly suited for scenarios where aggregated flow-level information provides sufficient detail for practical applications.

With this in mind, we introduce RouteNet-Gauss (hereafter referred to as RouteNet-G), an ML-based model trained on a representative dataset generated from a real-world testbed network to enhance network simulation. The dataset is created using a hardware testbed capable of dynamically producing diverse network configurations, including high-speed traffic profiles, varying topologies, and routing strategies. To address the lack of packet-level granularity, RouteNet-G leverages a configurable temporal granularity mechanism called \textit{Temporal Aggregated Performance Estimation} (TAPE). This mechanism aggregates detailed traffic information, such as individual packet traces, into temporal windows. By doing so, RouteNet-G provides flow-level metrics that maintain computational efficiency while capturing the temporal dynamics of flows.

While testbeds allow for high-fidelity training data, covering all possible network configurations—such as various routing schemes, topologies, or network sizes—remains challenging and resource-intensive, especially for large-scale networks. To address this challenge, RouteNet-G’s design aims to improve the model's ability to estimate performance metrics for a variety of scenarios, including those not explicitly represented in the training data, as detailed in Section~\ref{sec:overview_network}.

\begin{figure*}[t]
    \centering
    \includegraphics[width=\textwidth]{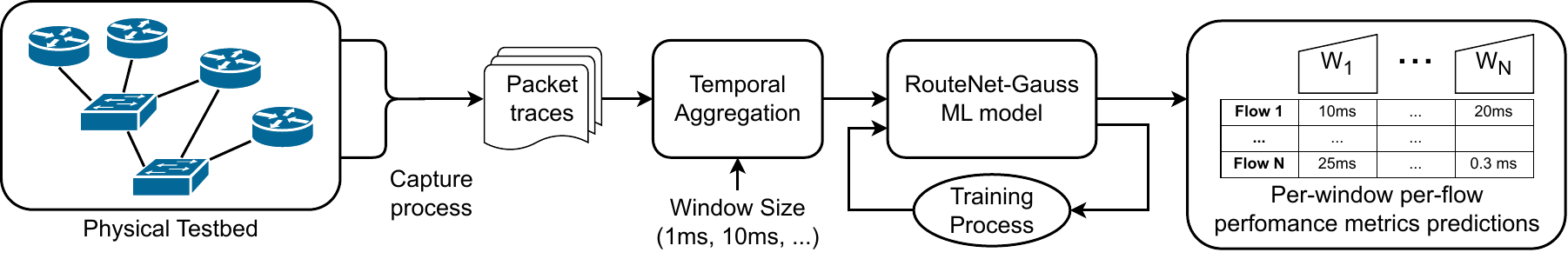}
    \caption{RouteNet-G's Workflow.}
    \label{fig:RouteNet-G}
\end{figure*}

RouteNet-G's workflow is shown in Figure~\ref{fig:RouteNet-G}. The model takes as input the network scenario (e.g., packet-level traffic, topology, routing) and returns sequences showing flow performance metrics over time. In our experiments, RouteNet-G shows remarkable accuracy when modeling scenarios not seen during training, with a Mean Absolute Percentage Error as low as 2.289\% in specific scenarios while significantly reducing inference time—up to 488 times faster than certain state-of-the-art solutions.


This paper is structured as follows: we first delve into the details of DES-based simulation and its limitations to motivate the design of RouteNet-G (Section \ref{sec:background}). We then present the design and implementation of RouteNet-G (Section~\ref{sec:solution}) and the testbed network (Section~\ref{sec:testbed}) used to generate the datasets for training the solution. We conclude by evaluating RouteNet-G and discussing its results (Section~\ref{sec:evaluation}), as well as its limitations and future directions (Section~\ref{sec:discussion}).

\section{Motivation}
\label{sec:background}



Discrete Event Simulation (DES) remains one of the most commonly used tools for network operators to model network scenarios. DES works by breaking down network activity into discrete ``events," such as packet generation, traversal through physical links, queuing, and eventual exit from the network. Each event is processed sequentially, updating the overall network state to simulate its behavior. This level of detail gives DES a high degree of accuracy, provided that the simulator faithfully updates the scenario's state to reflect real-world conditions.

Despite its strengths, the event-driven design of DES introduces certain limitations, particularly when it comes to parallelization. Simulators like ns-3~\cite{Riley2010} and OMNeT++~\cite{Varga2019} struggle with efficiently parallelizing event processing due to the inherent sequential nature of identifying and processing events. While these tools can divide network scenarios into smaller segments to process independently, events that span across multiple segments require inter-process communication. This overhead can significantly reduce overall performance and scalability~\cite{10.1145/3603269.3604844}.



To address these limitations, current state-of-the-art solutions focus on enhancing DES-based simulators with parallelization techniques (e.g., DONS~\cite{10.1145/3603269.3604844}) or by incorporating ML models to improve performance (e.g., DQN~\cite{10.1145/3544216.3544248}). While these approaches have made notable progress, DES-based simulations remain computationally expensive. Even with parallelization, the huge volume of events in large-scale network scenarios can make processing impractical. For instance, a single 1 Gbps link may process 250,000 500-byte packets per second, illustrating the immense computational demand of DES-based methods.

\subsection{DES's elevated computational cost}

\begin{figure}[!t]
    \centering
    \includegraphics[width=\columnwidth]{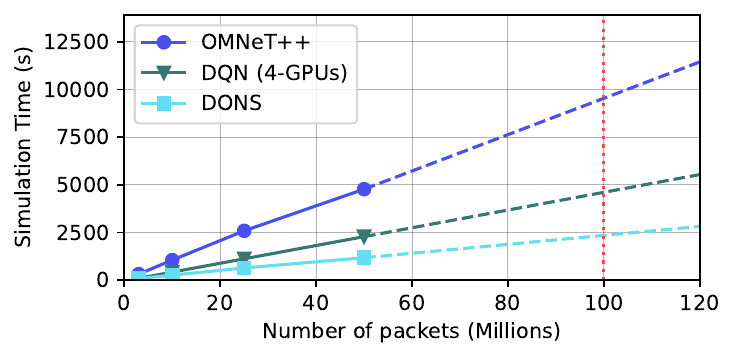}
    \caption{Relationship between the simulation cost and number of packets in the network scenario.}
    \label{fig:motivation_cost}
\end{figure}

To illustrate the high computational cost of DES, we refer to Figure~\ref{fig:motivation_cost}, showing the relationship between simulation time and the number of packets in a network scenario. This analysis is performed by using the same network setup while incrementally increasing the traffic volume. The dashed lines in the figure provide an approximation of the expected simulation times as the number of packets grows. Additionally, the dotted vertical line marks the number of packets transmitted by a 10 Gbps link operating at 80\% load for one second. This bandwidth reflects the minimum link capacities typically found in modern data centers~\cite{10.1145/2829988.2787472, Vahdat_2015}. The 80\% load represents a realistic peak for link utilization, often observed at an average network workload of 50\%~\cite{10.1145/3341302.3342085}.

As shown in the figure, both DONS and DQN —the latter utilizing 4 GPUs— successfully accelerate simulations compared to the legacy DES-based simulator OMNeT++. OMNeT++ takes approximately 80 minutes to simulate a network scenario with 5 million packets, while DQN completes the same task in 30 minutes. DONS, the fastest among the tested simulators, further reduces the simulation time to about 15 minutes. The figure also highlights the linear relationship between the volume of packets and simulation workload, underscoring the scalability challenges in network modeling.

However, despite these significant improvements in simulation speed, exemplified by DONS, even the fastest methods remain impractical for large-scale scenarios. For instance, simulating just one second of a typical data center network at a 10 Gbps link speed could still take nearly 2,500 seconds (over 40 minutes). While DONS represents a major step forward, the huge volume of packets in such scenarios poses a fundamental limitation, making DES-based methods difficult to use in time-sensitive applications.

\subsection{Inaccuracy of DES when compared to reality}

Another important but less frequently discussed challenge is the inaccuracies inherent in DES-based simulators. These tools are designed to model the behavior of network devices under specific scenarios. However, they often face limitations when the precise design and implementation details of commercial devices are unavailable, either due to proprietary restrictions or lack of documentation. This gap in information can lead to discrepancies between simulated and real-world results.

To demonstrate the extent of these inaccuracies, we conducted an experiment comparing the predictions of OMNeT++ with actual measurements obtained from running the same network scenarios on a physical testbed. The results from these traffic scenarios are further analyzed in Section~\ref{sec:evaluation} to assess the accuracy and reliability of the proposed solution.

\begin{table}[!t]
\centering
\resizebox{0.7\columnwidth}{!}{%
\begin{tabular}{lccc}
\toprule
     
    Dataset & MAPE & MAE ($\mu\text{s}$) & R\textsuperscript{2} \\
     \midrule
TREX Synthetic & 54.167\% & 80.682 & 0.716\\
TREX Multi-Burst & 46.911\% & 62.075 & 0.997\\
\bottomrule
\end{tabular}
}
\caption{
Differences in average packet delay per-flow per windows between OMNeT++~\cite{Varga2019} and the testbed.}
\label{tab:simulator_vs_testbed_raw}
\end{table}

The results are summarized in Table~\ref{tab:simulator_vs_testbed_raw}. The Mean Absolute Percentage Error (MAPE) and the Mean Absolute Error (MAE) were used to measure the discrepancies between the simulator's predictions and the actual testbed measurements. The results show that the simulator introduces a significant error, with MAPE values ranging from 46\% to 54\%, depending on the traffic distribution. In absolute terms, this corresponds to an average deviation of 62 to 80 $\mu\text{s}$ in the mean packet delay per flow within a time window. These findings align with the evaluation results presented later in Section~\ref{sec:evaluation}.


The R\textsuperscript{2} metric, which measures the correlation between predicted and actual values, provides additional insight into the simulator’s performance. For the first dataset, the R\textsuperscript{2} value was 0.716. While this value is not inherently poor, it does suggest that the simulator has difficulty fully replicating the testbed's behavior. The discrepancy appears to go beyond simple overestimation or underestimation, indicating that certain aspects of the network's dynamics are not being effectively captured within the simulation.


In short, network simulators face challenges in accurately modeling real-world networks. These limitations inevitably extend to current state-of-the-art ML models that rely on simulated data for training~\cite{10.1145/3603269.3604844, 10.1145/3452296.3472926, 10.1145/3544216.3544248, ferriolgalmés2022routenetfermi}, as the quality of an ML model is inherently tied to the quality and representativeness of the data it is trained on. Thus, any inaccuracies or gaps in the simulator’s ability to replicate network behavior can degrade the performance of these models when applied to real-world scenarios.

\section{RouteNet-Gauss}
\label{sec:solution}

In the following section, we propose RouteNet-G to address the limitations of existing simulation-based methods. By training on high-fidelity data from a physical testbed, RouteNet-G learns the intricate interactions between network components, enabling it to generalize effectively to unseen network scenarios. We first explain the intuition behind RouteNet-G and then describe its architecture, highlighting how it captures and leverages these interactions.

\subsection{RouteNet-Gauss's network decomposition}
\label{sec:overview_network}


RouteNet-G takes as input a network scenario, including the traffic, routing, and network configuration, and outputs per-flow performance metrics. This includes critical metrics such as delay, packet loss, and queue occupancy.

RouteNet-G's architecture works at a flow level. However, flows can be aggregated from individual flows to origin-destination (OD) flows —aggregating all flows that share the same routing path into one— depending on the desired level of granularity. Aggregating flows into OD-flows increases the model efficiency by reducing the number of elements to consider, at the cost of obtaining aggregated predictions. Furthermore, aggregation can be specifically useful when considering ``mice flows", flows with small payloads and flow completion times which form the overwhelming majority of traffic in modern networks \cite{kandula2009nature, jurkiewicz2021flow}. While individually they may be too small for their impact to be noticeable, their aggregate behavior can be more easily captured. In this paper, we will consider OD-flows, as they provide the level of granularity required for the use cases mentioned in Section~\ref{sec:intro}.


By training on high-fidelity testbed data, RouteNet-G is designed to model these scenarios accurately while maintaining the flexibility to generalize to unseen network topologies and configurations.
A straightforward approach —such as to model the entire network as a single neural network trained end-to-end— might seem appealing but introduces critical challenges. Specifically, such an architecture would be heavily biased toward the specific topologies and scenarios used during training, making it difficult to generalize to unseen or larger networks. To overcome this, RouteNet-G adopts a divide-and-conquer strategy: it models the individual elements in the network (i.e., queues, links, devices, and flows) and, more importantly, learns how these elements interact.

To do so, the simplest solution would consist on creating a graph that connects the different devices according to the network topology. However, this is not sufficiently detailed to capture the complexities of computer networks. For instance, a forwarding device that has several ports in it may or may not be affected by the status of the other devices, depending on the traffic present. Instead, RouteNet-G builds an extended graph that reflects the finer-grained interactions between network elements. The network topology and flow routing determine these interactions. For instance:

\begin{itemize}
    \item Devices in the network interact with queues at their interfaces.
    \item Queues are influenced by the traffic they process and interact with the links they serve.
    \item Flows interact with the sequence of queues and links they traverse, which affects their performance metrics such as delay and packet loss.
\end{itemize}

\begin{figure*}[t]
    \centering
    \includegraphics[width=\textwidth]{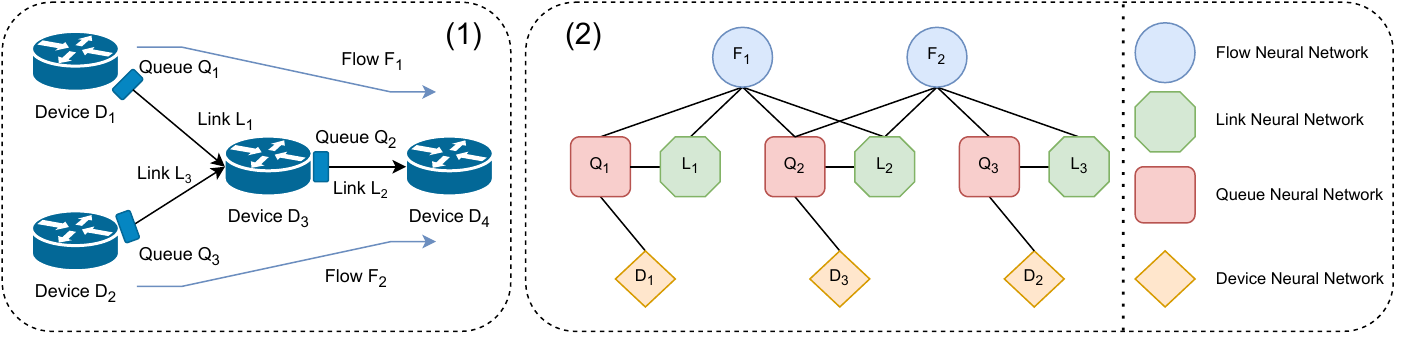}
    \caption{An example topology (1) and its RouteNet-G's expanded graph representation (2). The expanded graph represents the identified network elements and their interactions. Flows interact with the queues and links along their paths, links are influenced by the queues that inject them with traffic, and queues depend on the flows passing through them and the amount of shared resources they obtain from their network devices. According to these interactions, RouteNet-G dynamically builds its internal architecture using neural networks (NNs) that model the behavior of each element. These NNs are constructed during the training phase and used as ``building blocks" to build RouteNet-G's internal architecture at inference time. The same type of elements share the same NN. This approach allows RouteNet-G to work with any network topology it may face.}
    \label{fig:example}
\end{figure*}



This extended graph captures these interdependencies, allowing RouteNet-G to model the network’s behavior at a much deeper level. Figure~\ref{fig:example} illustrates this process with a simplified example. First, the raw network topology is used to generate an expanded graph (Figure~\ref{fig:example}(2)). For example, Flow 1 crosses Queue 1, Link 1, Queue 2, and Link 2. Each interaction is explicitly represented in the expanded graph. Second, each network element (e.g., queues, links, flows) is then represented by a corresponding neural network (NN), referred to as “building blocks.” These building blocks are shared across all elements of the same type. For instance, the same NN is used to model all queues, regardless of their position or the topology.



The use of shared NNs is central to RouteNet-G’s ability to generalize. By training these building blocks on diverse topologies, routing schemes, and traffic distributions, RouteNet-G learns the general principles of how network elements interact. This approach allows the model to dynamically adapt to new network scenarios without requiring retraining, even for topologies or configurations that are much larger or different from those seen during training. Evidence of this generalization capability is presented in Section~\ref{sec:generalization}.


RouteNet-G’s architecture is inspired by Message Passing Neural Networks (MPNN)\cite{10.5555/3305381.3305512}, a class of GNNs~\cite{scarselli2008graph}. MPNNs are specifically designed to handle graph-structured data and have demonstrated strong generalization capabilities across unseen graphs. RouteNet-G builds upon this foundation, extending and refining the architecture introduced in RouteNet-Fermi~\cite{ferriolgalmés2022routenetfermi} to incorporate the additional complexities of dynamic routing, flow-level granularity, and real-world testbed data.

\subsection{RouteNet-Gauss's temporal reasoning decomposition}

RouteNet-G incorporates the temporal dimension, striking a balance between computational efficiency and model expressiveness. Unlike traditional methods that either ignore the temporal aspect with aggregated predictions or process individual packets at high computational cost, RouteNet-G uses a middle-ground approach. It divides the network scenario into fixed-sized time windows and generates predictions for each window. This allows users to adjust the window size to suit their needs, balancing precision and efficiency. Although each window is processed independently, RouteNet-G reintroduces state dependencies using its internal state representations of network elements, encoding temporal information. This gives RouteNet-G the ability to handle scenarios with non-stationary traffic patterns.

For flows and links, RouteNet-G defines their initial state using measurable time-dependent metrics (e.g., current flow bandwidth and link load). This is not an option for queues and network devices, whose characteristics might be hard to measure in real-time (e.g., queue occupancy) or unaffected by time (e.g., buffer size). Instead, RouteNet-G uses these static descriptions to build the initial state in the first window. Then, the final internal states from one window become the initial states for the next, ensuring temporal continuity.

\subsection{RouteNet-Gauss's architecture}
\label{sec:algorithm}

\begin{algorithm}[!t]
\caption{RouteNet-Gauss}
\label{alg:RouteNet-G}
\begin{algorithmic}[1]
\Require {$\mathcal{F}, \mathcal{L}, \mathcal{Q}, \mathcal{N}, \boldsymbol{x}_f, \boldsymbol{x}_l, \boldsymbol{x}_q$}
\Ensure {$\hat{\boldsymbol{y}}_f$}

\ForAll {$q \in \mathcal{Q} $} $\boldsymbol{h}^{(0)}_{q} \gets E_{q}(\boldsymbol{x_q})$ \EndFor \label{alg:RouteNet-G-globalenc-start}
\ForAll {$d \in \mathcal{N} $} \label{line:while}
\State $M^0_d \gets \sum_{q \in d} \boldsymbol{h}^0_{q}$
    \State $\boldsymbol{h}^0_{d} \gets U_D(M^0_d; \boldsymbol{0})$
\EndFor \label{alg:RouteNet-G-globalenc-end}

\For{$t \gets 0$ to $T-1$} \label{alg:RouteNet-G-repeat}

    \ForAll {$f \in \mathcal{F}$} $\boldsymbol{h}^{(t)}_{f} \gets E_{f}(\boldsymbol{x^{(t)}_f})$ \EndFor \label{alg:RouteNet-G-localenc-start}
    \ForAll {$l \in \mathcal{L}$} $\boldsymbol{h}^{(t)}_{f} \gets E_{l}(\boldsymbol{x^{(t)}_l})$ \EndFor \label{alg:RouteNet-G-localenc-end}
    
    \State  $\boldsymbol{\hat{y}}^{(t)}_{f}, \{ \boldsymbol{h}^{(t+1)}_{q} | \forall q \in \mathcal{Q} \}, \{ \boldsymbol{h}^{(t+1)}_{d} | \forall d \in \mathcal{D} \} \gets MP \big( \{ \boldsymbol{h}^{(t)}_{f} | \forall f \in \mathcal{F} \}, \{ \boldsymbol{h}^{(t)}_{l} | \forall l \in \mathcal{L} \}, \{ \boldsymbol{h}^{(t)}_{q} | \forall q \in \mathcal{Q} \}, \{ \boldsymbol{h}^{(t)}_{d} | \forall d \in \mathcal{D} \}  \big)$ \label{alg:RouteNet-G-mpa}
\EndFor \label{alg:RouteNet-G-until}
\end{algorithmic}
\end{algorithm}

\begin{algorithm}[!t]
\caption{Message Passing Algorithm}
\label{alg:mpa}
\begin{algorithmic}[1]
\Require {$\boldsymbol{h}^0_f, \boldsymbol{h}^0_l, \boldsymbol{h}^0_q, \boldsymbol{h}^0_d;  \forall f \in \mathcal{F}, l \in \mathcal{L}, q \in \mathcal{Q}, d \in \mathcal{N} $}
\Ensure {$\hat{\boldsymbol{y}}_f, \boldsymbol{h}^T_q, \boldsymbol{h}^T_d; \forall f \in \mathcal{F}, q \in \mathcal{Q}, d \in \mathcal{N} $}

\For{$t \gets 1$ to $T$} \Comment{Message Passing Phase} \label{alg:mpa-repeat}
    \ForAll {$f \in \mathcal{F}$} \Comment{\footnotesize Message Passing on Flows} \label{alg:mpa-flows-start}
        \State $pos \gets 0$ \Comment{\footnotesize Curr. position in path}
        \State $\boldsymbol{h}^t_f \gets \boldsymbol{h}^{t-1}_f$
        \ForAll {$(l,q) \in f$}
            \State $\widetilde{m}^t_{f, pos}, \boldsymbol{h}^t_f \gets U_{F}(\boldsymbol{h}^{t-1}_{l} || \boldsymbol{h}^{t-1}_{q}; \boldsymbol{h}^t_f)$
            \State $pos \gets pos + 1$
        \EndFor
    \EndFor \label{alg:mpa-flows-end}
    \ForAll {$q \in \mathcal{Q}$} \Comment{\footnotesize Message Passing on Queues} \label{alg:mpa-queues-start}
        \State $M^t_{q} \gets \boldsymbol{h}^{(t-1)}_{\hat{Q}_d(q)} \bigparallel \sum_{(f, pos) \in \hat{Q}_{f}(q)}{\widetilde{m}^t_{(f, pos)}}$
        \State $\boldsymbol{h}^t_{q}, \emptyset \gets U_Q(M^t_{q}; \boldsymbol{h}^{t-1}_{q})$
        \State $\widetilde{m}^t_{q} \gets \boldsymbol{h}^t_{q}$
    \EndFor \label{alg:mpa-queues-end}
    \ForAll{$d \in \mathcal{D}$} \Comment{\footnotesize Message Passing on Devices} \label{alg:mpa-devices-start}
        \State $M^t_{d} \gets \sum_{q \in d}{\widetilde{m}^t_{q}}$ 
        \State $\boldsymbol{h}^t_d, \emptyset \gets U_D(M^t_{d}; \boldsymbol{h}^{t-1}_d)$ 
    \EndFor \label{alg:mpa-devices-end}
    \ForAll {$l \in \mathcal{L}$} \Comment{\footnotesize Message Passing on Links} \label{alg:mpa-links-start}
        \State $M^t_{l} \gets \widetilde{m}^t_{\hat{Q}_q(l)}$
        \State $\boldsymbol{h}^t_{l}, \emptyset \gets U_L(M^t_{l}; \boldsymbol{h}^{t-1}_{l})$
        \State $\widetilde{m}^t_{l} \gets \boldsymbol{h}^t_{l}$ 
    \EndFor \label{alg:mpa-links-end} 
\EndFor \label{alg:mpa-until}

\ForAll{ $f \in \mathcal{F}$} \Comment{Readout Phase} \label{alg:mpa-readout-start}
    \State $\boldsymbol{\hat{y}}_{f} \gets \sum_{\widetilde{m}^t_{f, pos} \in \widetilde{m}^t_f} R(\widetilde{m}^t_{f, pos})$
\EndFor \label{alg:mpa-readout-end}
\end{algorithmic}
\end{algorithm}

\renewcommand{\arraystretch}{1.5}
\begin{table}[t]
\centering
\resizebox{\columnwidth}{!}{%
\begin{tabular}{@{}lp{8.3cm}@{}}
\toprule
Symbol              & Definition / Description \\ \midrule
$\mathcal{F}$ & Set of flows in a network scenario \\
$\mathcal{L}$ & Set of links in a network scenario \\
$\mathcal{Q}$ & Set of queues in a network scenario \\
$\mathcal{N}$ & Set of network devices (nodes) in a network scenario \\
$E$ & Initial encoding function \\
$U$ & Update function of the MP phase \\
$R$ & Readout function \\
$T$ & Maximum iterations of the MP phase; a hyperparameter \\
$x_q$ & Input features of queue $q$ \\
$x^{(t)}_f$ & Input features of flow $f$ during window $t$ \\
$h_q^0$ & Initial encoding for queue $q$ \\
$h_q^t$ & Internal encoding for queue $q$ after iteration $t$ of the MP phase \\
$h_q^{(t)}$ & Internal encoding for queue $q$ after the window $t$ of scenario \\
$\widetilde{m}^t_{q}$ & Message generated by queue $q$ to be used by its neighbors at iteration $t$\\
\multirow{2}{*}{$\widetilde{m}^t_{f, pos}$} & Message generated by flow $f$ at position $pos$ of its routing path to be used by its neighbors at iteration $t$ \\
$M_q^t$ & Aggregation of messages for queue $q$ at iteration $t$ of the MP phase \\
$\hat{Q}_d(q)$ & Mapping function which returns the device the queue $q$ belongs to\\
\multirow{2}{*}{$\hat{Q}_f(q)$} & Mapping function which returns the flows which include the queue $q$ in its routing path\\
$\hat{Q}_q(l)$ & Mapping function which returns the exit queue of link $l$ \\
$\emptyset$ & Discarded output from an update function \\
$\hat{y}^{(t)}_{f}$ & Predicted performance metrics of flow $f$ at window $t$ \\
 \bottomrule
\end{tabular}%
}
\caption{Table of notations}
\label{tab:notations}
\end{table}
\renewcommand{\arraystretch}{1}

\begin{table}[]
\centering
\resizebox{\columnwidth}{!}{%
\begin{tabular}{@{}p{2cm}p{2.9cm}p{1.5cm}p{3.2cm}@{}}
\toprule
Element                  & Feature                   & Behavior over time & Notes                          \\ \midrule
\multirow{3}{*}{Flows}   & Average Load              & Variable           & In bits per second             \\ \cmidrule(l){2-4} 
                         & Rate of packets generated & Variable           & In packets per second       \\ \cmidrule(l){2-4} 
                         & Packet size               & Constant           & In bits                        \\ \midrule
Devices / Queues & Device type               & Constant           & Router, Switch or Endpoint     \\ \midrule
Link                     & Expected load             & Variable           & In \% of link's bandwidth \\ \bottomrule
\end{tabular}%
}
\caption{Extracted features of each element in the scenario.}
\label{tab:features}
\end{table}

We now proceed to explain RouteNet-G's architecture, which follows the same three-part structure as MPNNs: (1) encoding of the internal states for each element, (2) the Message Passing (MP) phase, where the internal states are updated with the information from neighboring elements, and (3) the readout phase, where the outputs of the MP phase are used to obtain the desired performance metrics. Algorithm~\ref{alg:RouteNet-G} represents RouteNet-G's architecture in pseudocode, while the MP and readout phases are further detailed in Algorithm~\ref{alg:mpa}.
The symbols used are also described in Table~\ref{tab:notations}.

First, the encoding of the internal states is shown in Algorithm~\ref{alg:RouteNet-G}, lines \ref{alg:RouteNet-G-globalenc-start}-\ref{alg:RouteNet-G-localenc-end}. The encoding takes the relevant input features for each network element, listed in Table~\ref{tab:features}, and encodes them using a Multi-Layer Perceptron (MLP). In the case of the initial states for network devices, $h^0_d$, the update function $U_D$ from the MP algorithm (to be discussed later) is used instead. The loop present in line 5 represents the temporal component of the algorithm. In it, the MP algorithm is executed for each window (line \ref{alg:RouteNet-G-until}).

The MP phase is presented in Algorithm~\ref{alg:mpa}, lines \ref{alg:mpa-repeat}-\ref{alg:mpa-until}. It is an iterative process, in which the internal states of each element are updated according to their interactions with each other. The outer loop represents the number of times the states of the network elements are updated. In its body, the ``building blocks" NNs introduced back in Section~\ref{sec:overview_network} are executed, here referred to as update functions $U$. Each function takes as input the relevant network elements and updates the current one. Note again that the same $U$ functions are applied for all network elements and across all windows, as otherwise, the model would not adapt to unseen topologies. All of these functions are modeled through the use of Gated Recurrent Unit (GRU) \cite{cho2014learning} cells. Finally, in the readout phase (lines \ref{alg:mpa-readout-start}-\ref{alg:mpa-readout-end}), the readout function $R$ takes as input the final flow states and returns the predicted performance metrics.
Prediction is done for each queue in the flow's routing path, and then added.
Function $R$ is implemented through an MLP.

\section{Testbed implementation}
\label{sec:testbed}
In the following section, we cover the specific testbed implementation used to run network scenarios.
The testbed, illustrated at Figure~\ref{fig:testbed}, is summarized as follows:

\begin{figure}[t]
    \centering
    \includegraphics[width=0.75\linewidth]{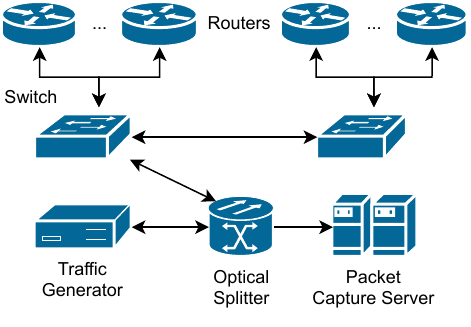}
    \caption{Diagram summarizing the testbed's structure.}
    \label{fig:testbed}
\end{figure}

\begin{itemize}
    \item The testbed comprises 8 Huawei NetEngine 8000 M1A routers connected to one of the two Huawei S5732-H48UM 2CC 5G Bundle switches.
    \item The switches act as the backbone, connecting the routers, and using VLANs to generate the desired topology. 
    \item A server generates the traffic, assigning a pool of addresses and VLANs per path, which is then forwarded through the network and back. TREX software is used to generate synthetic traffic.
    \item The traffic-generating server is connected to one of the switches. Traffic going through this connection is copied using the optical splitter and then sent to the packet capture server. The packet capture server runs custom-made software, which utilizes the DPDK framework and a Network Interface Controller, to record the hardware timestamp of each captured packet.
    \item Different link bandwidths are chosen to be representative of those in modern networks:
    \begin{itemize}
        \item Links between routers and switches are 1 Gbps each.
        \item Links between switches and the traffic generator and the packet capture server are 10 Gbps each.
        \item Both switches are connected through 2 $\times$ 40 Gbps links in trunk mode. 
    \end{itemize}
\end{itemize}



The main advantage of this setup is that the testbed can be reconfigured to create any topology formed by up to 8 nodes. Additionally, since the traffic is copied optically through the splitter, the packet capture process does not impact the performance of the testbed and its components, and it does not introduce any measurable delay to the recordings. 

\section{Evaluation}
\label{sec:evaluation}
To assess our claims, we have implemented RouteNet-G and trained it on network scenarios generated from the testbed network to answer the following questions:
\begin{itemize}
    \item Does RouteNet-G improve the computational cost compared to DES (\textit{Issue \#1})?
    \item Does RouteNet-G improve prediction accuracy compared to DES (\textit{Issue \#2})?
    \item Does RouteNet-G generalize to unseen network scenarios with differing amounts of traffic and network sizes?
    \item What is the impact of RouteNet-G's temporal aggregation?
\end{itemize}

All the following experiments use a server with an AMD Ryzen 9 3950X 16-Core 32 Threads 3.5 GHz CPU and 64 GiB of RAM. This server uses Ubuntu 22.04 LTS as its operating system. We implemented RouteNet-G using TensorFlow 2.11 \cite{199317}. The implementation details and source code are recollected both in Appendix~\ref{appendix:flownet_hyperparams} and in a public repository\footnote{https://github.com/BNN-UPC/Papers/wiki/RouteNet\_Gauss}.

\subsection{Evaluation of RouteNet-G's temporal cost}
\label{sec:eval_cost}

\begin{figure}[t]
    \centering
    \includegraphics[width=\columnwidth]{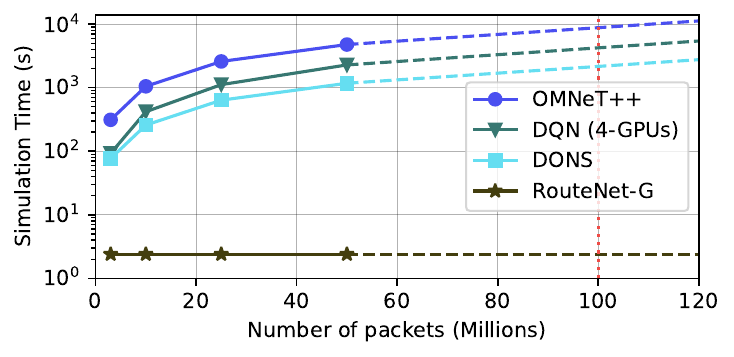}
    \caption{RouteNet-G's inference cost against DES. 
    Dashed lines indicate extrapolation from measured values, while the vertical dotted line indicates the expected number of packets seen per second in a modern datacenter link (10 Gbps link at 80\% load).
    }
    \label{fig:flownet_cost}
\end{figure}

First, we compare RouteNet-G's performance in evaluating network scenarios against current state-of-the-art methods. We specifically assess the inference time of the trained RouteNet-G model relative to other alternatives, including OMNeT++~\cite{Varga2019}, DONS~\cite{10.1145/3603269.3604844}, and DQN~\cite{10.1145/3544216.3544248}. Both RouteNet-G and the baseline methods are executed in a single thread, except for DQN, which utilizes 4 GPUs. The scenarios evaluated correspond to those illustrated in Figure~\ref{fig:motivation_cost}, back in Section~\ref{sec:background}. We analyze the inference time required by each method to process these scenarios and generate its corresponding predictions.

Figure~\ref{fig:flownet_cost} shows the cost of inference of RouteNet-G relative to the baselines. As shown, RouteNet-G can perform inference over the flow's behavior in the network scenarios in just 2.4 seconds, which at 50 million packets is 488x faster than the quickest alternative, DONS, which takes over 1000 seconds. Importantly, this duration remains consistent regardless of the number of packets in the scenario, unlike the baselines, whose inference time is influenced by the packet count and limits their applicability to larger scenarios.

Note that while RouteNet-G is not affected by the number of packets, other factors such as topology size and the number of windows in the network scenario may influence its performance (as shown in the following Section \ref{subsec:impact_of_temp_aggr}). Despite these factors, RouteNet-G substantially outperforms the next fastest baseline, indicating a significant reduction in computational complexity compared to current state-of-the-art solutions.

\subsection{Evaluation of RouteNet-G's accuracy}
\label{subsec:eval_accuracy}

Next, we evaluate RouteNet-G's accuracy when predicting flow performance metrics compared to OMNeT++~\cite{Varga2019}. As discussed earlier, current state-of-the-art solutions are either DES-based simulators (e.g., DONS, Parsimon) or ML models trained with samples generated through DES (e.g., DQN). Hence, using OMNeT++, an established DES-based simulator, acts as an upper bound for the accuracy achievable by the current network modeling solutions based on DES.
We also compare against our implementation of RouteNet-F, based on the public one available\footnote{https://github.com/BNN-UPC/RouteNet-Fermi}, and adapted and trained with the testbed datasets.


\begin{table}[t]
\centering
\resizebox{0.8\columnwidth}{!}{%
\begin{tabular}{@{}cllll@{}}
\toprule
\multicolumn{2}{c}{Traffic Distribution} & TREX-S              & TREX-MB             & RWPT                \\ \midrule   
Number        & Training    & $1376$                & $541$                 & $1682$                \\ \cmidrule(l){2-5} 
of            & Validation  & $274$                 & $108$                 & $336$                 \\ \cmidrule(l){2-5} 
network       & Test        & $173$                 & $51$                  & $172$                 \\ \cmidrule(l){2-5} 
scenarios     & Total       & $1823$                & $700$                 & $2190$                \\ \midrule
Number        & Training    & $12.64 \times 10^6$ & $8.75 \times 10^6$  & $12.47 \times 10^6$ \\ \cmidrule(l){2-5}       
of            & Validation  & $2.56 \times 10^6$  & $1.67 \times 10^6$  & $2.56 \times 10^6$  \\ \cmidrule(l){2-5}       
valid         & Test        & $1.67 \times 10^6$  & $8.75 \times 10^5$  & $1.24 \times 10^6$  \\ \cmidrule(l){2-5}       
predictions   & Total       & $16.87 \times 10^6$ & $11.30 \times 10^6$ & $16.27 \times 10^6$ \\ \bottomrule
\end{tabular}%
}
\caption{
Summary of datasets generated. Valid predictions are defined as the product between the number of flows and extracted windows per scenario.
}
\label{tab:datasets}
\end{table}

In each network scenario, we compare the per-flow and per-window metrics computed by RouteNet-G and OMNeT++. For instance, when predicting the average packet delay, we consider packets generated within the same flow and during the same temporal window. For each flow-window pair, we extract metrics such as the average, median, 90th, 95th, and 99th percentiles of the packet delays and jitter. While the model could also be used to predict other performance metrics, such as the packet loss rate, we decided to focus on the delay and jitter as they are two of the most commonly studied performance metrics in network modeling. The RouteNet-G model is trained using three datasets, each using a different traffic distribution, generated in our testbed:
\begin{enumerate}
    \item TREX Synthetic (TREX-S): packets are generated in regular-high frequency bursts, each under 1 ms.
    \item TREX Multi-burst (TREX-MB): packets are generated from synthetic multiple burst distributions, each defined by their intensity and spacing between bursts. The generated packets at a given instant are the sum of packets generated by each of the defined burst distributions.
    \item Real-World Packet Traces (RWPT): generated flows follow the traffic distribution from real-world traffic traces in the MAWI Working Group Traffic Archive \cite{10.5555/1267724.1267775}.
\end{enumerate}

The recorded network scenarios are split into training, validation, and test partitions.
Partitioning the datasets is common practice to later ensure that the model has not over-fitted to the training data.
Each of the samples in these datasets represents a network scenario, consisting of a combination of network topology, input traffic, routing configuration, and a collection of per-flow metrics.  Each dataset was generated across 11 distinct topologies, ranging from 5 to 8 nodes. Illustrations of these are recollected in Appendix~\ref{appendix:topologies}. Flow routings span paths traveling from 3 to 5 routers, resulting in an overall range of 12 to 28 hops in the link layer. Moreover, in each network scenario, the number of generated packets varies from 450 thousand to 16 million.
The number of scenarios per type and partition is indicated in Table~\ref{tab:datasets}.

\begin{figure}[!t]
    \centering
    \subfloat[TREX Synthetic.\label{fig:cbr_avg_delay_pdf}]{\includegraphics[width=0.47\textwidth]{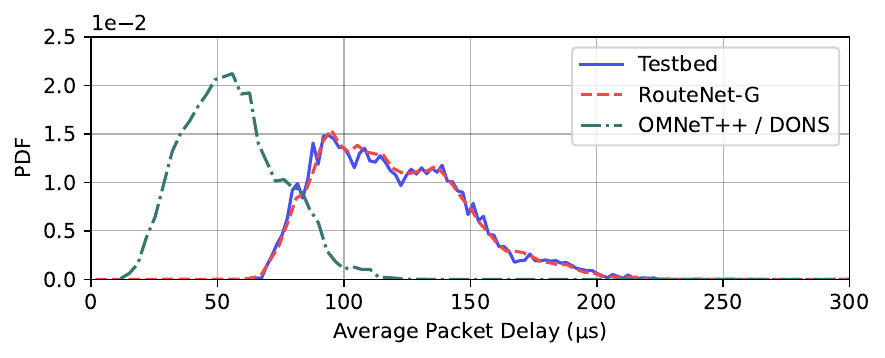}\hfill}\\
    \subfloat[TREX Multi-burst.\label{fig:cbr_mb_avg_delay_pdf}]{\includegraphics[width=0.47\textwidth]{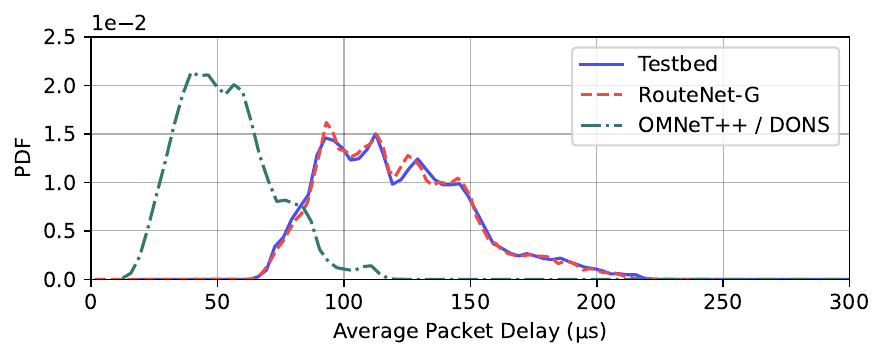}\hfill}    
    \caption{Probability Density Function of the predicted and measured average packet delay per each flow-window pair.}
    \label{fig:cdf_testbed_relative_err}
\end{figure}

\subsubsection{RouteNet-Gauss's accuracy in synthetic traffic}

Figure~\ref{fig:cdf_testbed_relative_err} illustrates the distribution of the measured average packet delay in each flow-window pair in the test partition of the synthetic datasets, as well as the distribution of the predicted values by both the OMNeT++ simulator and RouteNet-G.
In both datasets we can distinguish the lack of overlap between the measured average traffic delays in the testbed and OMNeT++'s predictions, indicating that DES-based simulation fails to accurately model the real behavior of the network. This is in contrast to RouteNet-G, which after training, replicates the traffic distribution of the testbed. We emphasize that in this scenario, OMNeT++ acts as an upper bound in the accuracy level of DES-based solutions, like DQN, and if these were to be plotted they are expected to perform even worse. 

\begin{table}[]
\centering
\resizebox{\columnwidth}{!}{%
\begin{tabular}{@{}cllccccc@{}}
\toprule
&       &           &       &           &           90th        & 95th          & 99th          \\
Dataset & Metric    & Model & Average   & Median & Percentile   & Percentile    & Percentile    \\
\midrule
\multirow{9}{*}{\rotatebox[origin=c]{90}{TREX-S}}
& \multirow{3}{*}{MAPE} 
& RouteNet-G & $\textbf{2.604\%}$ & $\textbf{2.790\%}$ & $\textbf{2.418\%}$ & $\textbf{2.591\%}$ & $\textbf{3.121\%}$ \\
& & RouteNet-F & $2.859\%$ & $3.007\%$ & $2.865\%$ & $2.911\%$ & $3.155\%$ \\
 & & OMNeT++ & $53.684\%$ & $54.325\%$ & $47.311\%$ & $45.609\%$ & $42.814\%$ \\ \cmidrule(lr){2-8}

& \multirow{3}{*}{MAE ($\mu\text{s}$)}  
& RouteNet-G & $\textbf{3.128}$ & $\textbf{3.305}$ & $\textbf{3.136}$ & $\textbf{3.442}$ & $\textbf{4.351}$ \\
& & RouteNet-F & $48.142$ & $48.287$ & $49.107$ & $49.349$ & $49.951$ \\
 & & OMNeT++ & $63.486$ & $63.714$ & $61.355$ & $60.859$ & $60.112$ \\ \cmidrule(lr){2-8}

& \multirow{3}{*}{R\textsuperscript{2}}   
& RouteNet-G & $\textbf{0.941}$ & $\textbf{0.939}$ & $\textbf{0.942}$ & $\textbf{0.935}$ & $\textbf{0.909}$ \\
& & RouteNet-F & $-0.001$ & $-0.001$ & $-0.001$ & $-0.001$ & $-0.001$ \\
 & & OMNeT++ & $-4.337$ & $-4.389$ & $-3.741$ & $-3.553$ & $-3.184$ \\ \midrule 

\multirow{9}{*}{\rotatebox[origin=c]{90}{TREX-MB}}
& \multirow{3}{*}{MAPE} 
& RouteNet-G & $\textbf{2.277\%}$ & $\textbf{2.480\%}$ & $\textbf{2.289\%}$ & $\textbf{2.529\%}$ & $\textbf{3.188\%}$ \\
& & RouteNet-F & $5.239\%$ & $5.126\%$ & $5.678\%$ & $5.994\%$ & $6.867\%$ \\
 & & OMNeT++ & $56.122\%$ & $57.353\%$ & $48.504\%$ & $46.355\%$ & $43.007\%$ \\ \cmidrule(lr){2-8}

& \multirow{3}{*}{MAE ($\mu\text{s}$)}  
& RouteNet-G & $\textbf{2.809}$ & $\textbf{3.006}$ & $\textbf{3.119}$ & $\textbf{3.523}$ & $\textbf{4.614}$ \\
& & RouteNet-F & $229.625$ & $228.217$ & $291.220$ & $300.024$ & $308.233$ \\
 & & OMNeT++ & $67.807$ & $68.646$ & $64.347$ & $63.354$ & $61.940$ \\ \cmidrule(lr){2-8}

& \multirow{3}{*}{R\textsuperscript{2}}   
& RouteNet-G & $\textbf{0.921}$ & $\textbf{0.924}$ & $\textbf{0.833}$ & $\textbf{0.826}$ & $\textbf{0.814}$ \\
& & RouteNet-F & $-0.007$ & $-0.006$ & $-0.008$ & $-0.008$ & $-0.008$ \\
 & & OMNeT++ & $-4.508$ & $-4.689$ & $-3.311$ & $-3.095$ & $-2.784$ \\ \bottomrule
\end{tabular}%
}
\caption{RouteNet-G's error when predicting packet delay across multiple traffic patterns.}
\label{tab:testbed_model_err_delay}
\end{table}

\begin{table}[]
\centering
\resizebox{\columnwidth}{!}{%
\begin{tabular}{@{}cllccccc@{}}
\toprule
&       &           &       &           &           90th        & 95th          & 99th  \\
Dataset & Metric    & Model & Average   & Median & Percentile   & Percentile    & Percentile \\ \midrule

\multirow{9}{*}{\rotatebox[origin=c]{90}{TREX-S}}
& \multirow{3}{*}{MAPE} 
& RouteNet-G & $9.447\%$ & $11.881\%$ & $9.347\%$ & $9.074\%$ & $9.609\%$ \\
& & RouteNet-F & $\textbf{8.476\%}$ & $\textbf{10.864\%}$ & $\textbf{8.547\%}$ & $\textbf{8.323\%}$ & $\textbf{9.050\%}$ \\
 & & OMNeT++ & $24.999\%$ & $27.549\%$ & $25.221\%$ & $24.172\%$ & $23.097\%$ \\ \cmidrule(lr){2-8}

& \multirow{3}{*}{MAE ($\mu\text{s}$)}  
& RouteNet-G & $0.792$ & $0.814$ & $1.610$ & $1.866$ & $2.558$ \\
& & RouteNet-F & $\textbf{0.703}$ & $\textbf{0.732}$ & $\textbf{1.447}$ & $\textbf{1.673}$ & $\textbf{2.324}$ \\
 & & OMNeT++ & $2.122$ & $1.919$ & $4.447$ & $5.118$ & $6.418$ \\ \cmidrule(lr){2-8}

& \multirow{3}{*}{R\textsuperscript{2}}   
& RouteNet-G & $0.758$ & $0.711$ & $0.754$ & $0.748$ & $0.700$ \\
& & RouteNet-F & $\textbf{0.802}$ & $\textbf{0.763}$ & $\textbf{0.793}$ & $\textbf{0.786}$ & $\textbf{0.734}$ \\
 & & OMNeT++ & $-0.584$ & $-0.509$ & $-0.799$ & $-0.702$ & $-0.482$ \\ \midrule 

\multirow{9}{*}{\rotatebox[origin=c]{90}{TREX-MB}}
& \multirow{3}{*}{MAPE} 
& RouteNet-G & $\textbf{10.711\%}$ & $\textbf{13.680\%}$ & $\textbf{10.934\%}$ & $\textbf{10.913\%}$ & $\textbf{11.824\%}$ \\
& & RouteNet-F & $17.993\%$ & $20.329\%$ & $21.256\%$ & $22.892\%$ & $28.479\%$ \\
 & & OMNeT++ & $37.435\%$ & $41.358\%$ & $39.311\%$ & $37.906\%$ & $35.792\%$ \\ \cmidrule(lr){2-8}

& \multirow{3}{*}{MAE ($\mu\text{s}$)}  
& RouteNet-G & $\textbf{0.967}$ & $\textbf{1.022}$ & $\textbf{1.949}$ & $\textbf{2.281}$ & $\textbf{3.112}$ \\
& & RouteNet-F & $1.452$ & $1.384$ & $2.955$ & $3.580$ & $5.361$ \\
 & & OMNeT++ & $3.244$ & $2.926$ & $7.005$ & $8.055$ & $9.818$ \\ \cmidrule(lr){2-8}

& \multirow{3}{*}{R\textsuperscript{2}}   
& RouteNet-G & $\textbf{0.529}$ & $\textbf{0.431}$ & $\textbf{0.625}$ & $\textbf{0.632}$ & $\textbf{0.633}$ \\
& & RouteNet-F & $0.154$ & $0.129$ & $0.229$ & $0.221$ & $0.160$ \\
 & & OMNeT++ & $-1.980$ & $-1.790$ & $-2.488$ & $-2.230$ & $-1.657$ \\
\bottomrule 
\end{tabular}%
}
\caption{RouteNet-G's error when predicting packet jitter across multiple traffic patterns.}
\label{tab:testbed_model_err_jitter}
\end{table}

We summarize the evaluation of the accuracy of both RouteNet-G and OMNeT++ across the remaining performance metrics in Table~\ref{tab:testbed_model_err_delay} and Table~\ref{tab:testbed_model_err_jitter}. The former shows the performance of RouteNet-G against OMNeT++ when predicting the packet delay, while the latter considers it when predicting the packet jitter. The accuracy of both alternatives is measured by the  Mean Absolute Percentage Error (MAPE), a relative error metric, the Mean Absolute Error (MAE), an absolute error metric, and the correlation coefficient (R\textsuperscript{2}), which measures the consistency of the predictions with reality.

Overall, the results show that RouteNet-G scores better than OMNeT++ across all three metrics, both while predicting the delay and the jitter. While predicting the delay, RouteNet-G's MAPE and MAE are $92.6\% - 95.9\%$ and $89.3\% - 95.1\%$ lower than OMNeT++, respectively. Specifically, RouteNet-G's MAE of $2.809 {\mu}s - 4.614 {\mu}s$ is similar to the expected transmission delay of packets in the dataset ($4 {\mu}s$), compared to OMNeT++'s MAE of $60.112 {\mu}s - 68.646 {\mu}s$. When predicting the jitter this difference is smaller but still significant, as RouteNet-G's MAPE is between $56.9\%$ and $72.2\%$ lower than OMNeT++'s MAPE. Furthermore, RouteNet-G's R\textsuperscript{2} scores across the board are higher, never dropping under 0.4, while OMNeT++'s scores are always negative. 

\begin{figure}[t]
    \centering
    \includegraphics[width=\columnwidth]{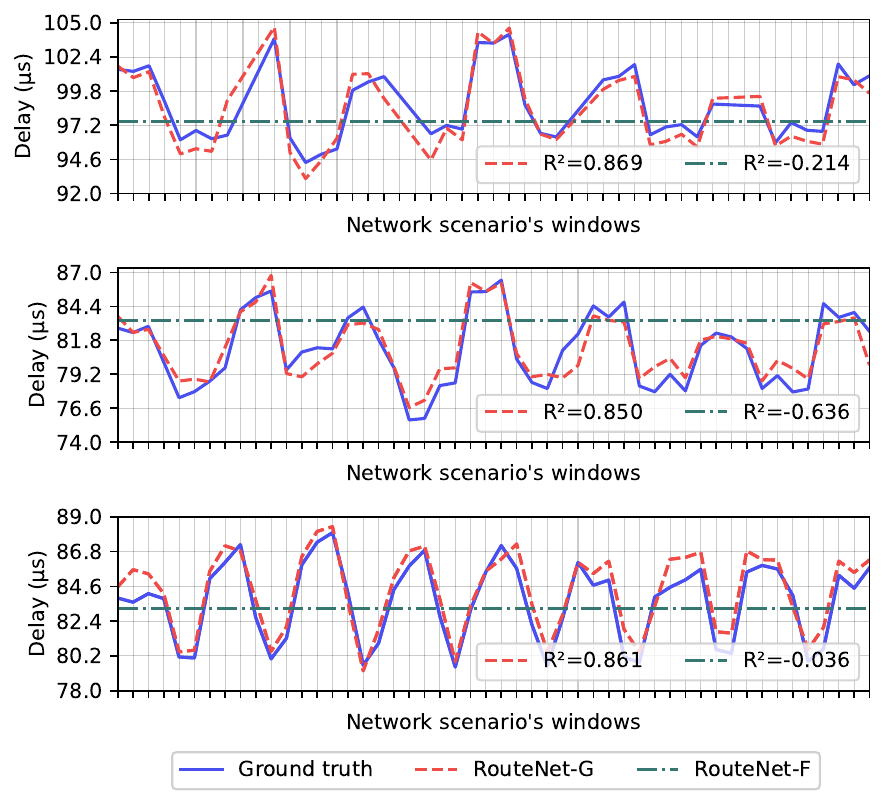}
    \caption{
    Qualitative comparison between RouteNet-G and RouteNet-F predictions for three example TREX-MB flows.
    }
    \label{fig:fermi_vs_flownet_testbed_delay}
\end{figure}

Comparing RouteNet-G's performance against RouteNet-F's generally shows the former outperforms the latter. In the TREX-S, the performance between the two is similar, with RouteNet-G with more accurate delay predictions and RouteNet-F with more accurate jitter predictions. However, in the TREX-MB dataset, which considers a more complex traffic profile, RouteNet-G outperforms RouteNet-F in both predicting delay and jitter, and across both metrics.

Overall, RouteNet-G's temporal granularity allows it to be more accurate. This is illustrated in Figure~\ref{fig:fermi_vs_flownet_testbed_delay}, showing the delay predictions made by RouteNet-G and RouteNet-F in three random flows from the TREX-MB dataset. While RouteNet-F only predicts the average delay across the entirety of each of the flow's duration, RouteNet-G does reflect how the delay evolves over time. This increased predictive granularity becomes more relevant the more complex the underlying traffic patterns are, as evidenced by the widening performance between both versions of RouteNet in TREX-MB and, later, in the RWPT dataset.

\subsubsection{RouteNet-Gauss's accuracy in real traffic data}
\label{subsec:real_data_eval}

\begin{figure}
    \centering
    \includegraphics[width=\linewidth]{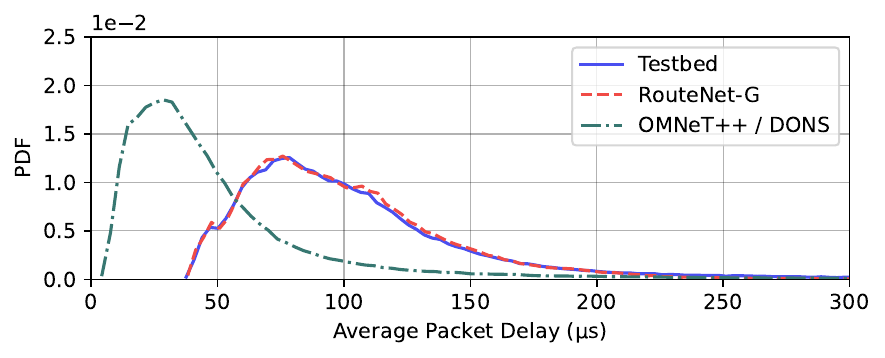}
    \caption{
    Probability Density Function of the predicted and measured average packet delay per each flow-window pair in the RWPT dataset.
    }
    \label{fig:mawi_pdf}
\end{figure}

After proving RouteNet-G's effectiveness with synthetic traffic, we next study how RouteNet-G performs when working with real traffic traces. Table~\ref{tab:testbed_model_err_mawi} shows both RouteNet-G's and OMNeT++'s accuracy when predicting the delay and jitter in the RWPT dataset. Figure~\ref{fig:mawi_pdf} compares the distributions of the predicted packet mean packet delay by RouteNet-G and OMNeT++ against the ground truth.

\begin{table}[]
\centering
\resizebox{\columnwidth}{!}{%
\begin{tabular}{@{}cllccccc@{}}
\toprule
Target & Error &         &       &          &  90th & 95th & 99th  \\
Metric & Metric    & Model & Average   & Median & Percentile   & Percentile    & Percentile \\ \midrule

\multirow{9}{*}{\rotatebox[origin=c]{90}{Delay}}
& \multirow{3}{*}{MAPE}

& RouteNet-G & $\textbf{12.990\%}$ & $\textbf{8.904\%}$ & $\textbf{18.439\%}$ & $\textbf{21.638\%}$ & $30.016\%$ \\ 
& & RouteNet-F & $34.894\%$ & $10.999\%$ & $38.863\%$ & $64.659\%$ & $166.097\%$ \\
& & OMNeT++ & $55.808\%$ & $64.239\%$ 
& $39.786\%$ & $34.860\%$ & $\textbf{28.182\%}$ \\ \cmidrule(lr){2-8}

& \multirow{3}{*}{MAE ($\mu\text{s}$)}  
 & RouteNet-G & $\textbf{44.564}$ & $\textbf{24.515}$ & $126.583$ & $173.643$ & $270.073$ \\ 
 & & RouteNet-F & $64.943$ & $26.477$ & $148.715$ & $236.762$ & $700.745$ \\
& & OMNeT++ & $57.350$ & $55.541$ & $\textbf{62.892}$ & $\textbf{68.590}$ & $\textbf{91.492}$ \\ \cmidrule(lr){2-8}

& \multirow{3}{*}{R\textsuperscript{2}}  
& RouteNet-G & $0.080$ & $0.046$ & $0.054$ & $0.055$ & $0.029$ \\ 
& & RouteNet-F & $0.138$ & $0.031$ & $0.148$ & $0.158$ & $-1.924$ \\
 & & OMNeT++ & $\textbf{0.939}$ & $\textbf{0.931}$ & $\textbf{0.972}$ & $\textbf{0.952}$ & $\textbf{0.811}$ \\ \midrule

\multirow{9}{*}{\rotatebox[origin=c]{90}{Jitter}}
& \multirow{3}{*}{MAPE} 
 & RouteNet-G & $12.773\%$ & $18.356\%$ & $13.184\%$ & $12.198\%$ & $15.587\%$ \\ 
& & RouteNet-F & $14.938\%$ & $22.531\%$ & $15.418\%$ & $14.142\%$ & $20.768\%$ \\
& & OMNeT++ & $\textbf{9.399\%}$ & $\textbf{16.056\%}$ & $\textbf{8.505\%}$ & $\textbf{8.816\%}$ & $\textbf{11.973\%}$ \\ \cmidrule(lr){2-8}

& \multirow{3}{*}{MAE ($\mu\text{s}$)}  
& RouteNet-G & $2.126$ & $1.335$ & $5.439$ & $7.435$ & $19.427$ \\ 
& & RouteNet-F & $2.371$ & $1.591$ & $6.018$ & $8.036$ & $21.424$ \\
 & & OMNeT++ & $\textbf{1.443}$ & $\textbf{1.090}$ & $\textbf{3.385}$ & $\textbf{4.959}$ & $\textbf{13.113}$ \\ \cmidrule(lr){2-8}

& \multirow{3}{*}{R\textsuperscript{2}}   
& RouteNet-G & $0.628$ & $0.735$ & $0.513$ & $0.437$ & $0.320$ \\ 
& & RouteNet-F & $0.602$ & $0.667$ & $0.473$ & $0.429$ & $0.347$ \\
& & OMNeT++ & $\textbf{0.843}$ & $\textbf{0.840}$ & $\textbf{0.825}$ & $\textbf{0.788}$ & $\textbf{0.662}$ \\ \bottomrule

\end{tabular}%
}
\caption{RouteNet-G's error when working with RWPT.}
\label{tab:testbed_model_err_mawi}
\end{table}

Overall, comparing higher error metrics when predicting real traffic to those when predicting synthetic traffic shows that the former is harder to predict. RouteNet-G and OMNeT++ perform similarly, with the specifics depending on the target performance metric. While predicting the delay, RouteNet-G outperforms DES when predicting the average and median delay, reducing the MAPE by $86\%$ and the MAE by $55.87\%$. When predicting the high percentile delays, both methods offer similar performance, with RouteNet-G scoring better MAPE values but worse MAE values. Overall, we also see RouteNet-G's R\textsuperscript{2} scores are generally close to 0, while OMNeT++'s are high. Then, when predicting the jitter, OMNeT++ slightly yet consistently performs better than RouteNet-G: in terms of MAPE, for example, we see that OMNeT++ improves RouteNet-G's score by up to $35\%$ (when predicting the 90th percentile jitter, from $13.184\%$ MAPE to $8.505\%$).

Altogether, the results show that RouteNet-G and OMNeT++ perform similarly, with differences mainly depending on the performance metric to predict. That being said, we must take into account that RouteNet-G can achieve these at a significantly lower computational cost.
Note that OMNeT++, as a DES-based simulator, is expected to obtain at least the same or better accuracy as ML models trained with simulated data (e.g., DQN, Parsimon). This suggests that a comparison between RouteNet-G and one of these alternatives would be even more favorable towards RouteNet-G. 
The results also show RouteNet-G regularly outperforming RouteNet-F across all performance metrics.
Overall, the results show that the benefits of RouteNet-G and using a testbed network also extend to captured traffic traces.

\subsection{RouteNet-Gauss's generalization}
\label{sec:generalization}

One of the key aspects of RouteNet-G is its generalization capabilities when working with unseen scenarios. In the previous sections, we showed RouteNet-G's ability to model different traffic distributions and generalize across topologies. In this section, we focus on proving RouteNet-G's generalization towards larger, unseen topologies than those seen during training. This property is essential to RouteNet-G's ability to work on production networks, as it would be prohibitively expensive to build a testbed network that size. Similarly, our testbed network is limited to up to 8 node configurations, so we exceptionally use simulated samples to generate the training and evaluation samples to test RouteNet-G's generalization capabilities. We argue that the source of the data does not condition RouteNet-G's ability to generalize.

\begin{figure}
    \centering
    \includegraphics[width=\linewidth]{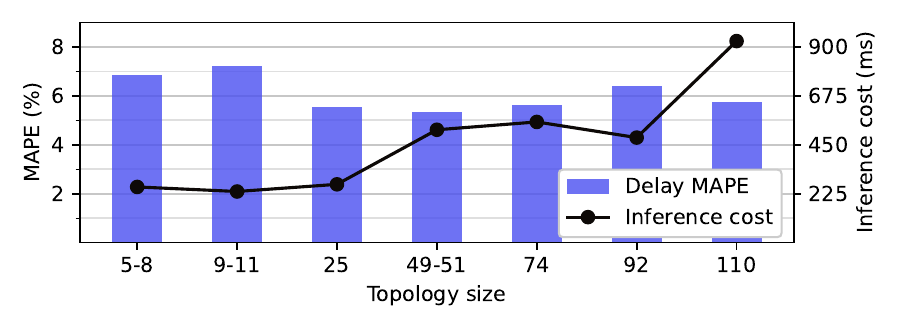}
    \caption{RouteNet-G's MAPE and inference cost at different sized topologies.}
    \label{fig:generalization}
\end{figure}

In this section, the RouteNet-G model is trained using network scenarios with topologies ranging from 5-8 nodes and then evaluated in unseen topologies with up to 110 nodes gathered from the Topology Zoo repository~\cite{6027859}. Samples followed the TREX-MB traffic distribution. The model was tasked to predict the average packet delay per-packet and per-window. The results are illustrated in Figure~\ref{fig:generalization}. The bar plot indicates the MAPE obtained in each range of topologies, while the line plot shows the mean sample inference cost.

The results show that the MAPE remains stable across all topologies, ranging between 5-7\%, confirming RouteNet-G's ability to generalize to larger topologies. On the other side, the inference cost does increase slightly with topology size: by performing linear regression on the trendline, we obtain that on average, the inference cost increases by 5.4ms per node in the original topology with $R^2$ of 0.90. Nevertheless, the small increase, combined with the overall small inference cost compared to the DES-based solutions as explored back in Section~\ref{sec:eval_cost}, suggests RouteNet-G's ability to properly work in larger topologies under a reasonable cost. 
Later in Section~\ref{sec:discussion_generalization}, we further discuss how this is the case.

\subsection{Impact of RouteNet-G's temporal aggregation}
\label{subsec:impact_of_temp_aggr}

In this section, we explore how RouteNet-G's temporal aggregation affects its predictions, its accuracy, and its costs.
As discussed earlier, RouteNet-G temporal aggregation allows for more granular and precise predictions. However, this also introduces a new hyperparameter, the window size, that defines the level of precision.

\begin{figure}[t]
    \centering
    \includegraphics[width=\columnwidth]{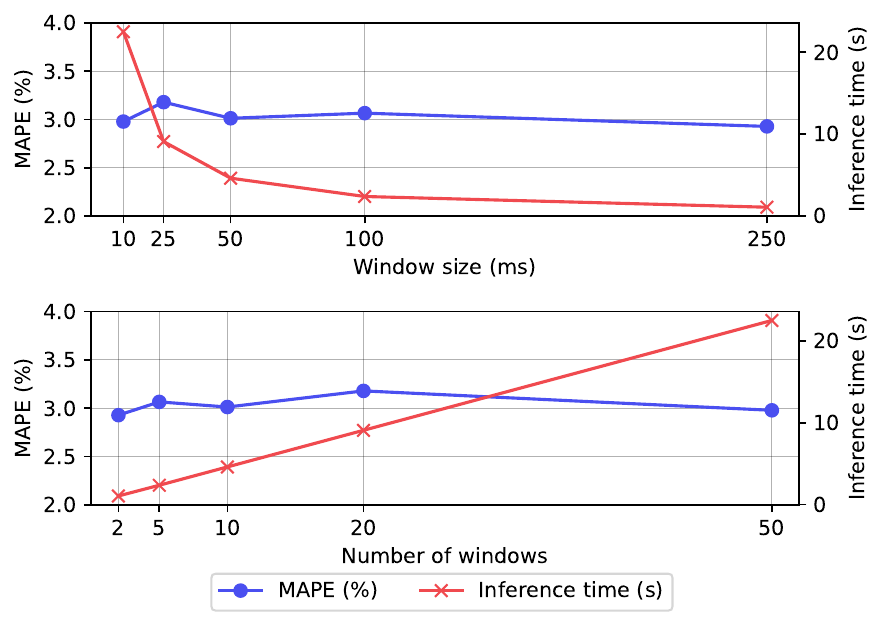}
    \caption{Impact of RouteNet-G window length over accuracy and cost.}
    \label{fig:window_size_vs_accuracy_cost}
\end{figure}

In Figure~\ref{fig:window_size_vs_accuracy_cost} we examine how changing the window size affects both RouteNet-G's accuracy and inference cost. On one hand, changing the window size does not impact the model's accuracy, showing its effectiveness when doing either detailed or summarized predictions. However, decreasing the window size results in a higher number of windows and, consequently, increasing the amount of predictions to perform. So, while the number of packets does not impact RouteNet-G's cost, as proved in Section~\ref{sec:eval_cost}, reducing the window size increases the computational demand.



\section{Discussion and limitations}
\label{sec:discussion}
This paper proposes a novel approach to improve the accuracy of network modeling by leveraging a testbed for training. By grounding the model in real-world data, we aim to address the limitations of simulation-based methods while enabling it to generalize to production networks, which are often much larger and more complex than the testbed used for training. This balance between high-fidelity data and scalability is at the core of RouteNet-G's design.

\subsection{Generalization to larger networks}
\label{sec:discussion_generalization}
One of the standout aspects of RouteNet-G is its ability to generalize to unseen network topologies, including those significantly larger than the ones encountered during training.

As discussed in Section~\ref{sec:overview_network}, RouteNet-G processes a network topology and its traffic by decomposing the problem into an expanded graph representation. This representation allows the model
to learn how various network components—such as links, queues, routers, switches, and flows—interact during training.
These interactions, such as if the buffer is empty, partially filled, or full, appear and influence network performance independently of network topology and size. Hence, by grounding the prediction's logic behind these interactions, a trained RouteNet-G model can remain sound even in unseen topologies.
When facing an unseen, larger network, the resulting expanded graph representation will be similar to those seen during training, as the interactions that drive its behavior remain the same. Consequently, RouteNet-G remains accurate, even in
networks of orders of magnitude larger than those in the training set, as evidenced in Section~\ref{sec:generalization}.

While the interactions between the network elements remain the same even in larger topologies, there are still two factors that may influence the resulting performance metrics, which are essentially dependent on network size: the overall larger (link) capacities and the longer routing paths. RouteNet-G's design also addresses these challenges to ensure its generalization to larger networks.


\subsubsection{Larger link capacities and aggregated traffic}
In larger networks, link capacities and the amount of traffic traversing them often scale alongside the network size. RouteNet-G addresses this by using scale-invariant features, such as link load, which capture relative metrics rather than absolute values. For instance, instead of representing raw link capacities, the model uses the ratio of utilized capacity to total capacity, enabling consistent reasoning regardless of network size.


\subsubsection{Path lengths in larger networks}
Another consideration in larger networks is the length of paths that flows follow. Since RouteNet-G treats paths as sequences of network elements (links and queues), it relies on seeing paths of similar lengths during training. This is mitigated by the fact that, in many computer networks, path lengths scale logarithmically with the number of nodes $N$ ($log(N)$ or even $\frac{log(N)}{log(log(N))}$)~\cite{CHEN20081405}. The connectivity of the network and the routing protocol used also influence this behavior.

By tuning the testbed to create paths within these expected ranges, we can emulate the conditions of larger networks effectively. For example, in the experiments evaluating RouteNet-G's generalization to larger topologies (Section~\ref{sec:generalization}), the 5-8 node topologies used for training included flow routing paths up to 5 routers long. On the other hand, in the 110 node topology, we measured routing paths up to 6 routers long.

In addition to this, RouteNet-G’s readout architecture (Section~\ref{sec:algorithm}) is designed to be robust against different path lengths by exploiting the additive nature of performance metrics. Performance metrics can be defined by the sum of each queue's impact within the flow path. Rather than predicting the performance metric directly, RouteNet-G's readout function first predicts the queue occupancy for each queue, uses them to derive the desired metric, and adds the results together to obtain the final flow-level prediction. As a result, the process naturally integrates the path length's impact in its prediction.

Finally, all of this discussion assumes we have a properly trained RouteNet-G model. Like any other ML method, the training set for RouteNet-G must be diverse to cover as many interactions present in production networks as possible. Hence, correct data sampling from the testbed is a must, or the prediction accuracy of the GNN will be diminished.

\subsection{Handling arbitrary traffic distributions}
One way RouteNet-G differs from its predecessors, like RouteNet-Fermi, is how it encodes traffic information. Unlike them, RouteNet-G neither makes assumptions about the flow's traffic distribution nor uses parameters of such distributions as part of its input. Instead, RouteNet-G describes flows utilizing the evolution of the traffic bandwidth and packet rate over the scenario duration.

This allows RouteNet-G to learn realistic non-parametric traffic flows, such as the RWPT dataset (Section~\ref{subsec:eval_accuracy}). However, unlike the features used to characterize link capacity, these are scale-dependent. Consequently, the model performs well with traffic distributions similar to those seen during training but may struggle if drastically different.
This highlights the importance of having a well-sampled dataset to train the model, including scenarios with high and low loads.
Addressing this limitation would require replacing scale-dependent features with invariant alternatives in future work.

    
    



\subsection{
Adapting RouteNet-G to TCP
}

So far, we have evaluated RouteNet-G using UDP traffic. However, the majority of internet traffic is regulated through congestion control algorithms, such as TCP. Including TCP samples in RouteNet-G's training is required for the model to support it. However, to the extent of our knowledge and given the differences between TCP and UDP, this may still be insufficient. In this section, we discuss how RouteNet-G can be adapted to support TCP and the challenges it may face.

First, flow aggregation most likely will be required. Back in Section~\ref{sec:overview_network}, we discussed aggregating flows into OD-flows. While RouteNet-G can support modeling individual flows, aggregating flows into OD-flows improves the computational cost and facilitates the modeling of ``mice flows" while still offering enough granularity for the proposed use-cases. For example, in \cite{jurkiewicz2021flow}, $58\%$ of recorded internet flows involved only 4 or fewer packets. These transmissions are so small that they are completed before being affected by TCP’s flow and congestion control mechanisms. Such small flows also tend to be more volatile, as losing and retransmitting a packet has a bigger impact on its average delay relative to larger flows. 

Conversely, by aggregating ``mice flows", the individual impact TCP has on each flow is softened as we consider the aggregate behavior. Instead, the aggregate behavior of these flows mostly depends on the link capacities and queue states. RouteNet-G has already proven able to learn interactions between traffic flows and the current network state, so it is reasonable to consider that it can successfully model aggregated TCP traffic. Moreover, it is well-known that TCP's congestion control algorithm directly depends on the amount of bandwidth available. Similar to how RouteNet-G learns when queues become congested and how it affects a flow's delay, it is also sound that RouteNet-G can learn and understand how much available bandwidth each flow has and, consequently, how TCP's congestion control algorithm will react.

However, in its current form, RouteNet-G lacks sufficient information to fully predict TCP's traffic behavior. For example, TCP flow control's receiver window for each host will depend on its capacity and current load, values currently unknown. To maximize RouteNet-G's accuracy when predicting TCP traffic, these values must be part of RouteNet-G's input when obtaining the initial queue or node encodings of the affected hosts. Another challenge is the heterogeneity present in TCP flavors in modern networks: RouteNet-G must be informed which implementation is currently in use in a given network scenario. A straightforward way to do so is to encode this information through one-hot encoding and include it as input to either the TCP flows' encoding or the affected queues and hosts. If the TCP flavor includes configuration parameters (e.g., ECN, RED), they must also be included alongside the TCP flavor. Finally, the training set must be expanded to cover both different TCP flavors and valid host configurations (e.g., transmission window sizes). RouteNet-G's hyperparameters may also require adjustments (e.g., prediction window sizes).

In summary, we believe RouteNet-G can be adapted to support TCP without fundamental changes to its architecture. These changes involve adding additional input features related to the present TCP implementations and hosts, and aggregating traffic into OD-flows. We leave these adaptations, as well as their evaluation, as future work.

\subsection{
Generalization to unseen hardware
}

While RouteNet-G demonstrates strong generalization to various network sizes, it was ultimately trained on a specific set of routers and switches. Like any ML-based model, RouteNet-G relies on its training data to learn the network interactions. Consequently, RouteNet-G can only be expected to generalize to hardware similar to that seen during training. We define the similarity of network devices by their specifications (e.g., memory size) and data plane forwarding technologies they implement, such as the QoS policies they support. 

For deployment in networks with devices from different manufacturers, additional training with data from those devices would likely be necessary. On the one hand, capturing traffic from production networks may prove challenging in some aspects, and testing multiple hardware devices in a testbed network may be expensive. On the other hand, hardware updates within the same network are relatively infrequent. Hence, updates to RouteNet-G will also be equally rare. Additionally, due to the properties of ML models, RouteNet-G does not require observing all possible combinations of available hardware and traffic distributions to achieve accuracy. Consequently, increasing the number of hardware devices supported does not result in a combinatorial explosion in the number of network scenarios sampled.

In addition to this, there are approaches that mitigate the strain of adapting RouteNet-G to unseen hardware during training. First, we can exploit previously trained RouteNet-G models to accelerate and decrease the cost of training models on new hardware. This is achieved by using transfer learning and domain adaptation techniques (e.g., fine-tuning~\cite{10.5555/2969033.2969197}). These techniques can extract trained knowledge from the previously trained RouteNet-G model and bootstrap the training of a new model better suited for a specific use case or hardware platform. This process requires substantially fewer measurements to train the model than training from scratch. The viability of this approach within the context of network performance modeling has been recently studied~\cite {Hattori2025MetaMetrics}.

Second, we may improve RouteNet-G's generalization to unseen hardware by parameterizing hardware features and introducing them as model inputs whenever possible. For example, QoS support can be handled by identifying each port's QoS algorithm (e.g., weighted queuing) and its parameters, as was done in RouteNet-Fermi~\cite{ferriolgalmés2022routenetfermi}. This would allow a RouteNet-G model to generalize to a similar router that also implements QoS support, even with differences in their configuration (e.g., different weighted queuing weights).


\subsection{Impact of the temporal component and TAPE}

One of the key innovations of RouteNet-G is its ability to work with different temporal granularities (TAPE). While previous RouteNet versions, including RouteNet-Fermi~\cite{ferriolgalmés2022routenetfermi}, assume stationary traffic, RouteNet-G relaxes the assumption by only expecting stationary traffic within the same window. As a result, it can model non-stationary traffic by analyzing it within time windows of an appropriate size—that is, the more volatile the traffic, the smaller the windows.

Similarly, TAPE also makes RouteNet-G suitable to model time-varying distributions (i.e., traffic patterns whose behavior changes over time). Modeling time-varying traffic is a complex, open problem~\cite{liu2016approximations, whitt2018time}. RouteNet-G addresses this complexity by basing its predictions on the current window's measurable state of the network. By focusing on the present, it can adapt to changes in behavior as it can in non-stationary traffic distributions. An example of this is traffic following the TREX-MB distribution, which alternates between periods of activity and inactivity lasting several seconds each.

That said, this flexibility comes with some trade-offs. First, as explored in Section~\ref{subsec:impact_of_temp_aggr}, smaller windows result in increased inference cost, albeit still under the cost of running the same scenario in DES-based alternatives. Second, RouteNet-G must be retrained if users need to work with different time scales. For example, predictions made using 10ms windows cannot simply be extrapolated to a 1ms window without retraining. While these limitations do not diminish the current contributions, they highlight an area for future improvement—such as enabling dynamic adaptation to multiple time scales.

\section{Related works}


The field of network modeling via simulation has a long trajectory within network research, with early works such as REAL~\cite{keshav1988real} dating back to 1988. Nevertheless, thanks to new technologies and programming paradigms, such as Machine Learning, the field is continuing to evolve even as of today.

Originally, network modeling research fell into one of two fields: simulation and analytical models. Analytical models such as Queuing Theory and Network Calculus \cite{robertazzi2000computer, arashloo2023formal}, benefited from quick inference time relative to simulation. However, these models fell out of favor due to their stringent assumptions, such as using over-simplistic distributions (e.g. Poisson distribution) to describe network traffic, which is measured to be auto-correlated with a heavy tail \cite{xu2018experience, popoola2017empirical, ArfeenRoleWeibull2019}.

Arguably, Discrete Event Simulation (DES) is one of the most widespread tools to simulate any kind of network to obtain performance estimations. Paramount examples are \mbox{NS-3}~\cite{Riley2010} and OMNeT++~\cite{Varga2019}, two network simulators with a large community, and the possibility to add plugins to expand their functionalities. As we mentioned, these simulators offer packet-level visibility, albeit with a high computational cost~\cite{netsim_how_fast2003}. 
Therefore, we can find multiple proposals that aim to accelerate these simulators without losing packet-level visibility. On one hand, there is a line of research dedicated to the parallelization and acceleration of DES engines, such as DONS~\cite{10.1145/3603269.3604844}, that makes use of Data-Oriented Design to increase the computational efficiency of the simulation engine. However, as we commented, DONS struggles in scenarios where the number of packets becomes too large to model individually. Alternatively, there is Parsimon~\cite{ZhaoParsimon2023}, which works by simulating each link separately, aggregating packet delays, and then calculating the path delay as a sum of its parts. While more efficient, it no longer achieves packet-level granularity and accuracy with respect to other DES-based simulators. Furthermore, it is purposefully designed to measure the tail flow-completion time of flows (90th-99th percentiles), while other solutions may be more general-purpose.

On the other hand, there is a growing body of research that leverages ML tools to replace parts of the DES engine to reduce computation time. Here we can find proposals such as
MimicNet~\cite{10.1145/3452296.3472926}, which takes advantage of the symmetry of data center networks to train a neural network that models a single top-level branch, and then extrapolates its results to the entire network. However, by design, it is restricted to Fat Tree topologies. Consequently, this inspired DQN~\cite{10.1145/3544216.3544248}, which uses data from a network simulator to train a Transformer neural network to predict the delay of batches of packets along the network. While not restricted to a specific topology, it is still limited by its computational complexity.

Furthermore, some approaches completely remove simulation during the inference process. These ``end-to-end" models use ML models to predict the desired performance metrics directly. A clear example is the previous iterations of RouteNet, such as RouteNet-Fermi~\cite{ferriolgalmés2022routenetfermi}, which trains an end-to-end graph neural network to provide average flow-level performance estimations. While it has the lowest computational cost, it can only form individual predictions for each flow over the entire scenario, lacking the expressiveness of both RouteNet-G and other solutions in the state-of-the-art.

Lastly, there have been some proposals trying to address the inaccuracies present in simulation. The Pantheon benchmark \cite{216073} provides improved emulators built from accurate network data derived from a variety of testbeds. There is also iBox \cite{10.1145/3508026}, which builds a time series model using the curated data from Pantheon. While the use of testbed-generated data is similar to our approach, both iBox and Pantheon focus on single-path prediction and the impact of congestion control, while RouteNet-G studies the entire network and the interactions between the network elements and packet flows.

\section{Conclusions}
In summary, this paper introduces a novel approach to enhance the cost-effectiveness and accuracy of network modeling by replacing DES with real network hardware. Through the utilization of a testbed network instead of a DES-based simulator, we address both computational cost concerns and inaccuracies arising from idealized scenario assumptions.

Using this approach we propose RouteNet-Gauss, a modular ML model trained using samples obtained from an implemented testbed network, whose architecture can adapt to the scenario's topology and routings, and which supports TAPE to balance computational cost and output expressiveness. Experimental results demonstrate that once trained, the RouteNet-G model achieves inference speeds two orders of magnitude faster than the quickest alternatives in the current state-of-the-art. Additionally, RouteNet-G exhibits superior accuracy in predicting flow performance metrics compared to DES-based simulators. Furthermore, our findings highlight RouteNet-G's flexibility when adjusting its computational complexity by varying the level of aggregation, at the expense of reduced output granularity, showcasing its versatility in addressing diverse network modeling requirements.

\section*{Acknowledgments}
This publication is part of the Spanish I+D+i project TRAINER-A (ref.PID2020-118011GB-C21), funded by MCIN/ AEI/10.13039/501100011033. This work is also partially funded by the Catalan Institution for Research and Advanced Studies (ICREA).
This work was also supported by the CHISTERA grant CHIST-ERA-22-SPiDDS-02 corresponding to the GRAPHS4SEC project (reference nº PCI2023-145974-2) funded by the Agencia Estatal de Investigación through the PCI 2023 call.
Carlos Güemes is funded by the AGAUR-FI ajuts (Grant Ref. 2023 F-1 00083) Joan Oró of the Secretariat of Universities and Research of the Department of Research and Universities of the Generalitat of Catalonia and the European Social Plus Fund.
Jordi Paillisse is funded by NextGen EU, Ministry of Universities and Recovery, Transformation and Resilience Plan, through a call from UPC (Grant Ref. 2022 UPC-MSC-93871).


\appendices
\section{RouteNet-Gauss hyperparameters}
\label{appendix:flownet_hyperparams}
\begin{table}[ht]
\centering
\resizebox{\columnwidth}{!}{%
\begin{tabular}{@{}p{2.2cm}p{4cm}p{3cm}@{}}
\toprule
\multirow{2}{2.2cm}{Pre-processing}  & Window size $(\Delta t)$          & 100 ms                                            \\ \cmidrule(l){2-3} 
                                     & Feature normalization          & Z-scores                                            \\ \midrule
\multirow{2}{*}{MP Algorithm}        & Dimensionality Embedded States  & 32                                                \\ \cmidrule(l){2-3} 
                                     & Number iterations               & 8                                                 \\ \midrule
\multirow{7}{2.5cm}{Training Parameters} & Maximum number of epochs        & 300                                               \\ \cmidrule(l){2-3} 
                                     & Number of steps per epoch       & 500                                               \\ \cmidrule(l){2-3} 
                                     & Optimizer                       & Adam                                              \\ \cmidrule(l){2-3} 
                                     & Learning rate $\alpha$          & 0.001                                             \\ \cmidrule(l){2-3}
                              & \multirow{2}{4cm}{Reduce Learning Rate Callback} & Patience = 10 \\ \cmidrule(l){3-3}
                              &                                                & Factor = 0.5  \\ \cmidrule(l){2-3} 
                                     & Loss function                   & MAPE                                              \\ \bottomrule
\end{tabular}%
}
\caption{Hyperparameters for the RouteNet-G Model.}
\label{tab:hyperparams}
\end{table}

The RouteNet-G model has been implemented using Python 3.10.12 and Tensorflow 2.11.1. The hyperparameters and their values are described in Table~\ref{tab:hyperparams}. Z-score normalization is applied specifically to the flow's average bandwidth and packet rate. The "Reduce Learning Rate Callback" is a function that triggers if the validation loss does not improve after 10 epochs, which then results in the learning rate being cut by half. The final epoch used is the one that minimized the MAPE of the validation set during training.

\section{Illustrations of generated topologies}
\label{appendix:topologies}

\begin{figure*}[ht]
    \centering
    \subfloat[]{\includegraphics[width=0.25\textwidth]{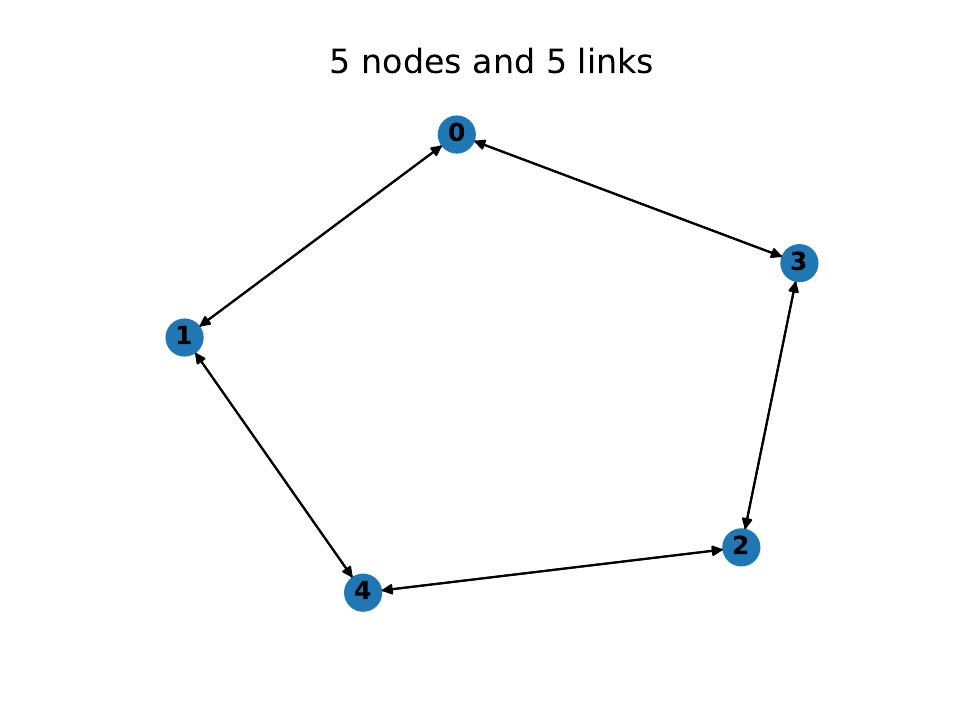}\hfill}
    \subfloat[]{\includegraphics[width=0.25\textwidth]{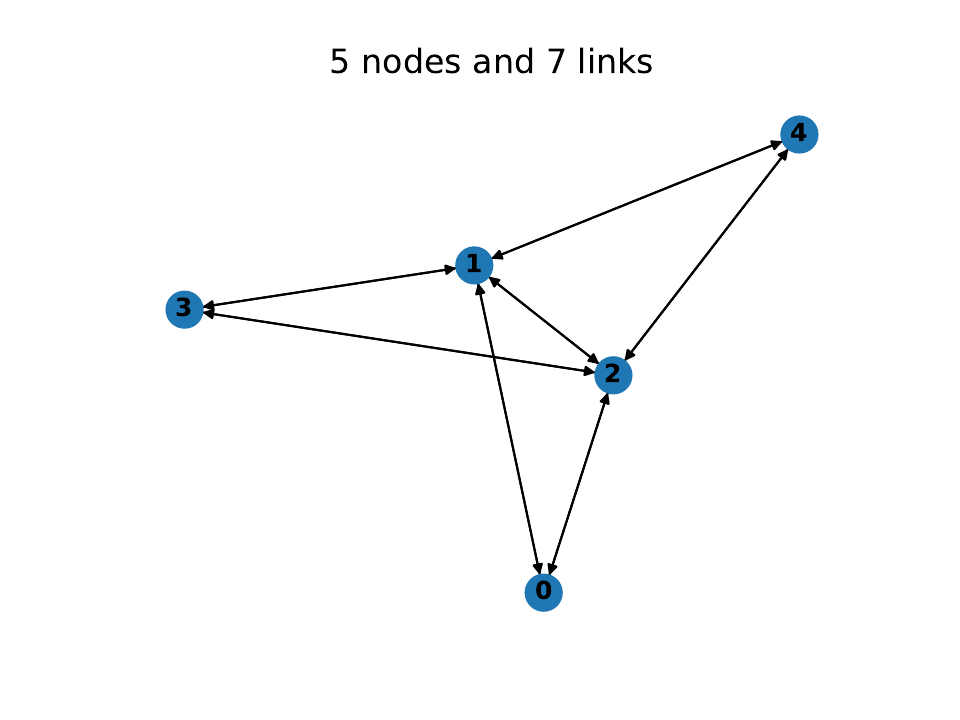}\hfill}
    \subfloat[]{\includegraphics[width=0.25\textwidth]{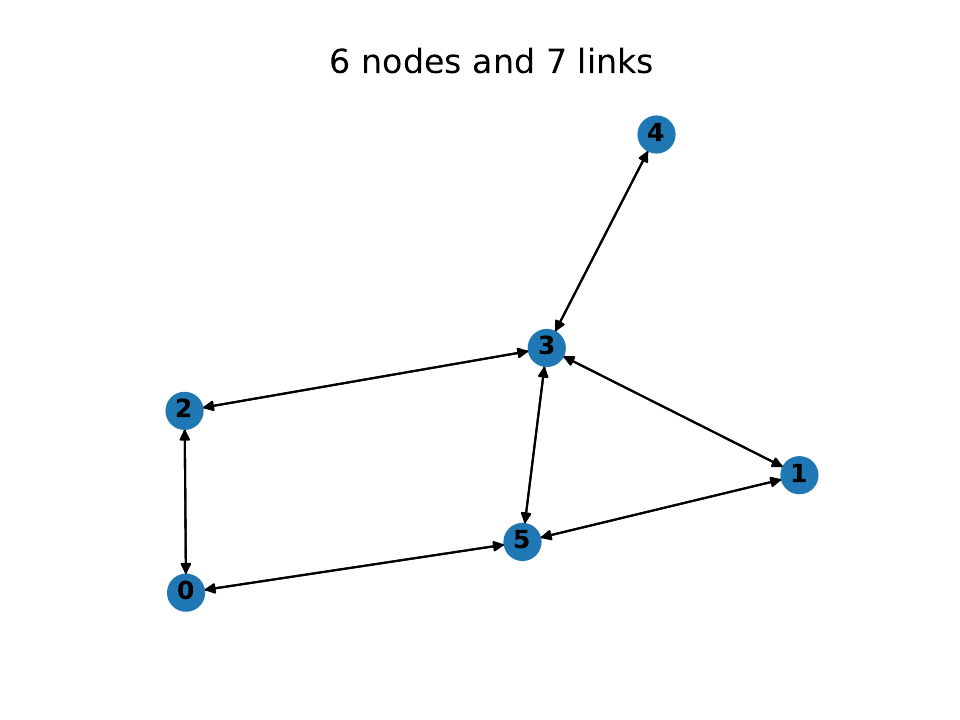}\hfill} 
    \subfloat[]{\includegraphics[width=0.25\textwidth]{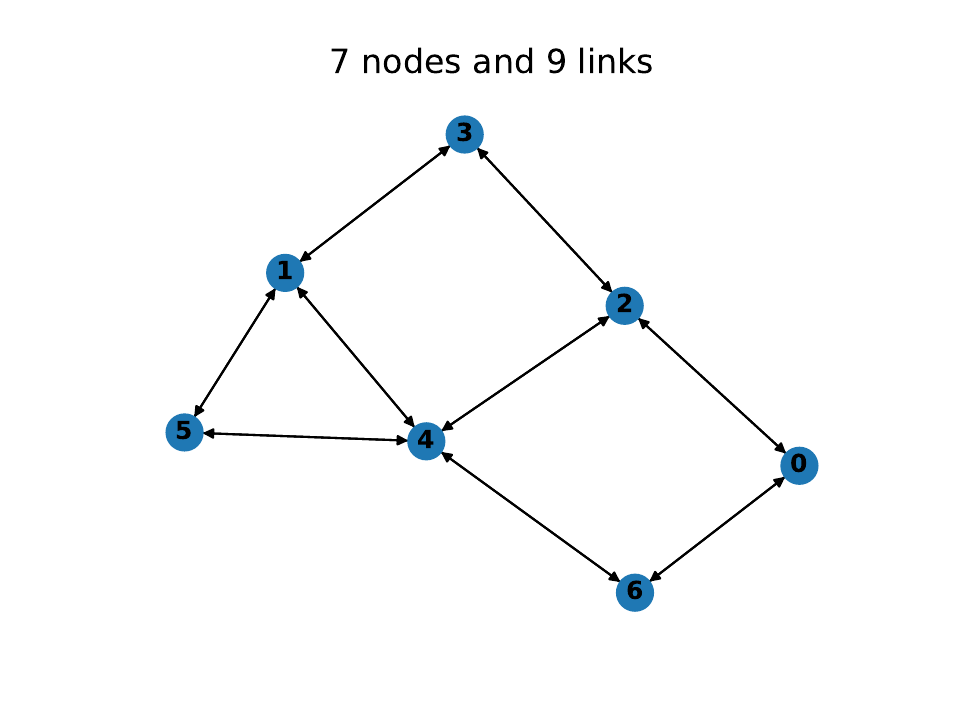}\hfill} \\
    \subfloat[]{\includegraphics[width=0.25\textwidth]{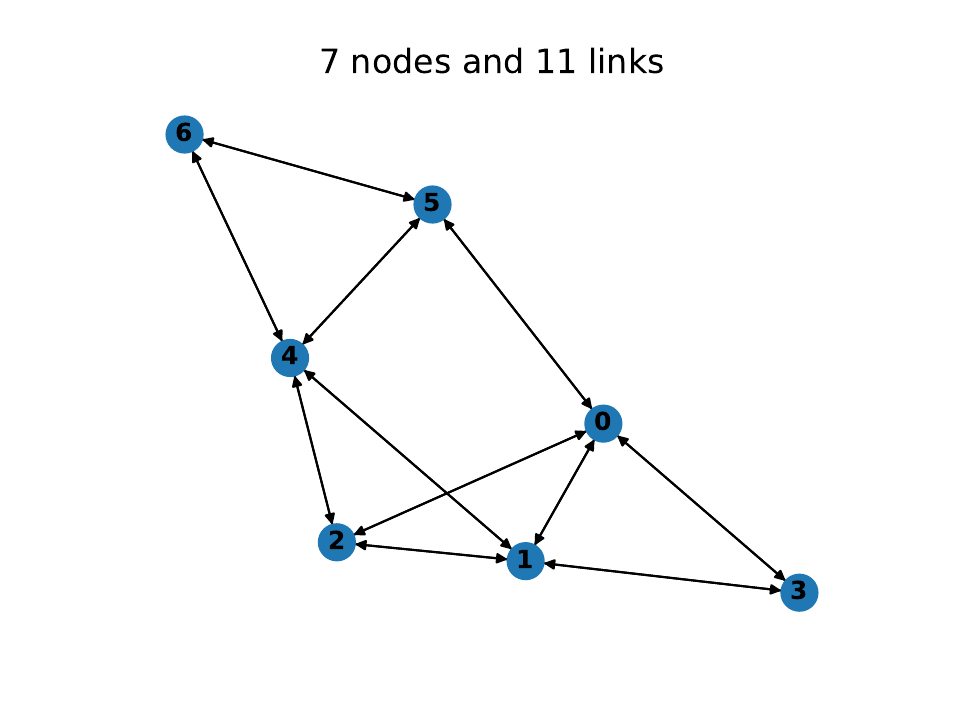}\hfill}
    \subfloat[]{\includegraphics[width=0.25\textwidth]{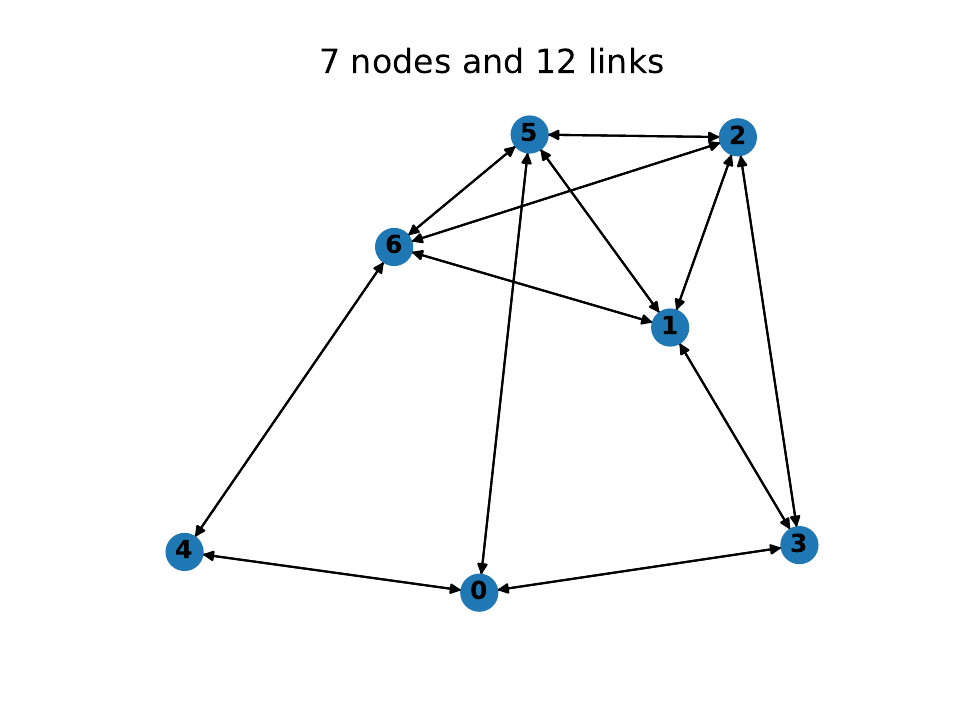}\hfill}
    \subfloat[]{\includegraphics[width=0.25\textwidth]{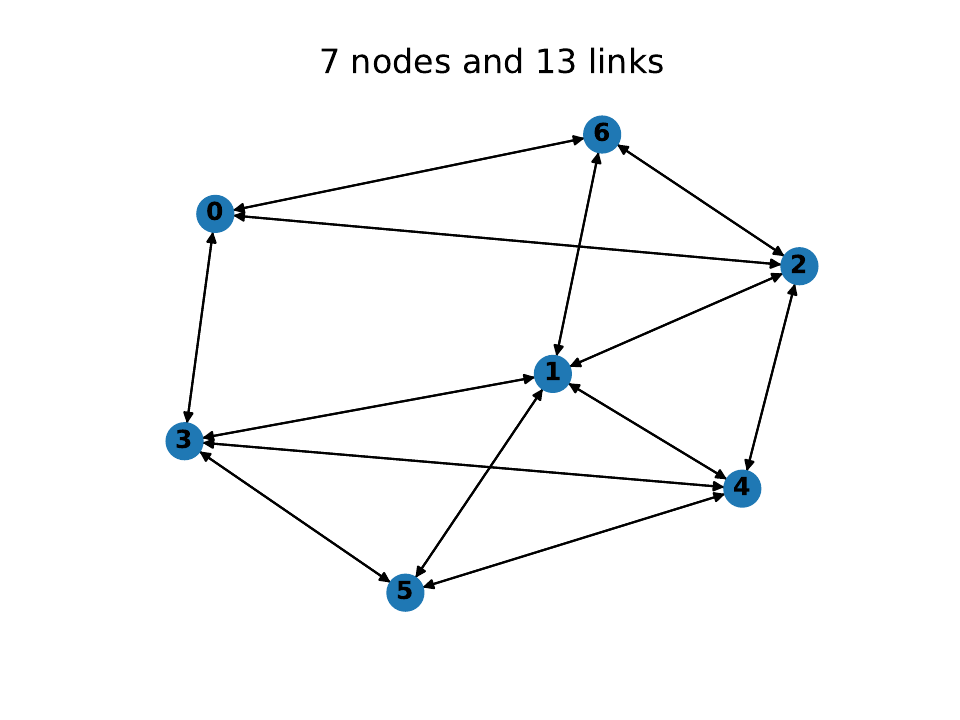}\hfill}
    \subfloat[]{\includegraphics[width=0.25\textwidth]{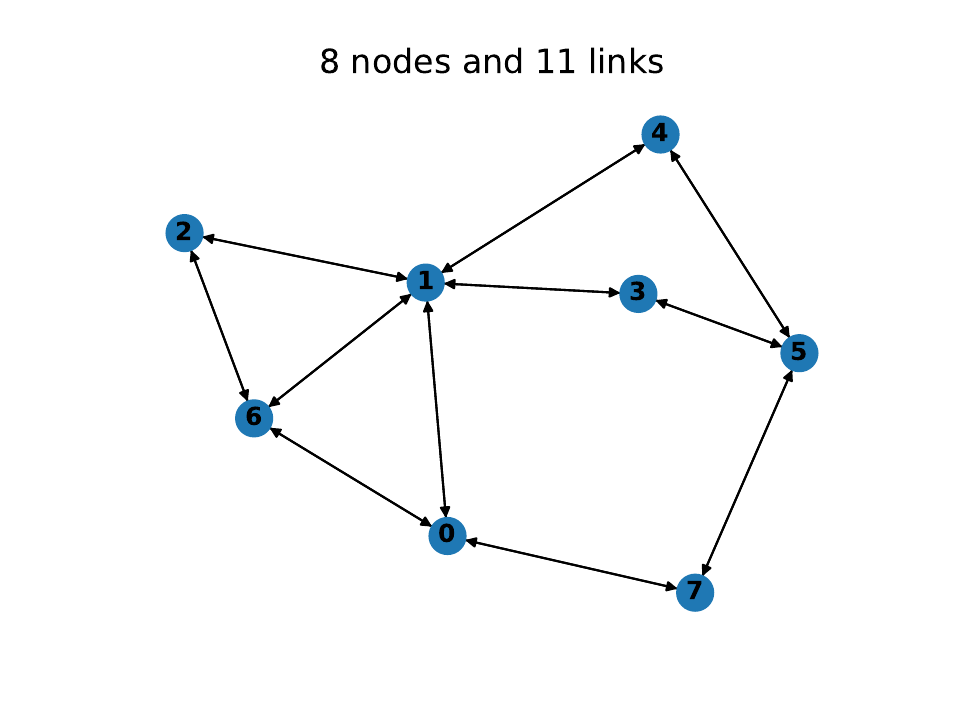}\hfill} \\
    \subfloat[]{\includegraphics[width=0.25\textwidth]{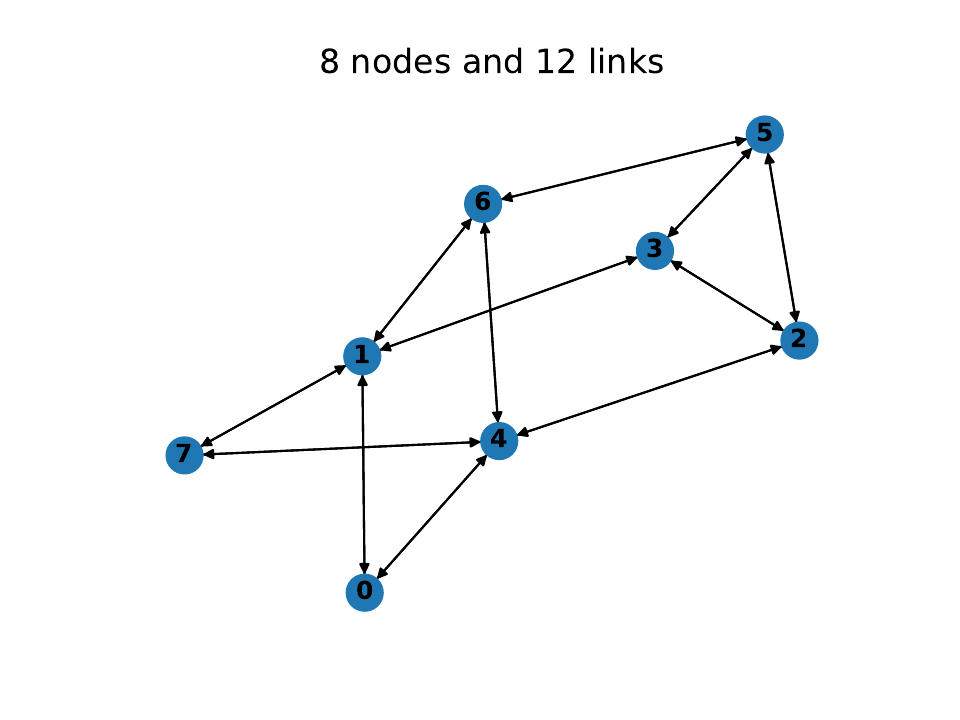}\hfill}
    \subfloat[]{\includegraphics[width=0.25\textwidth]{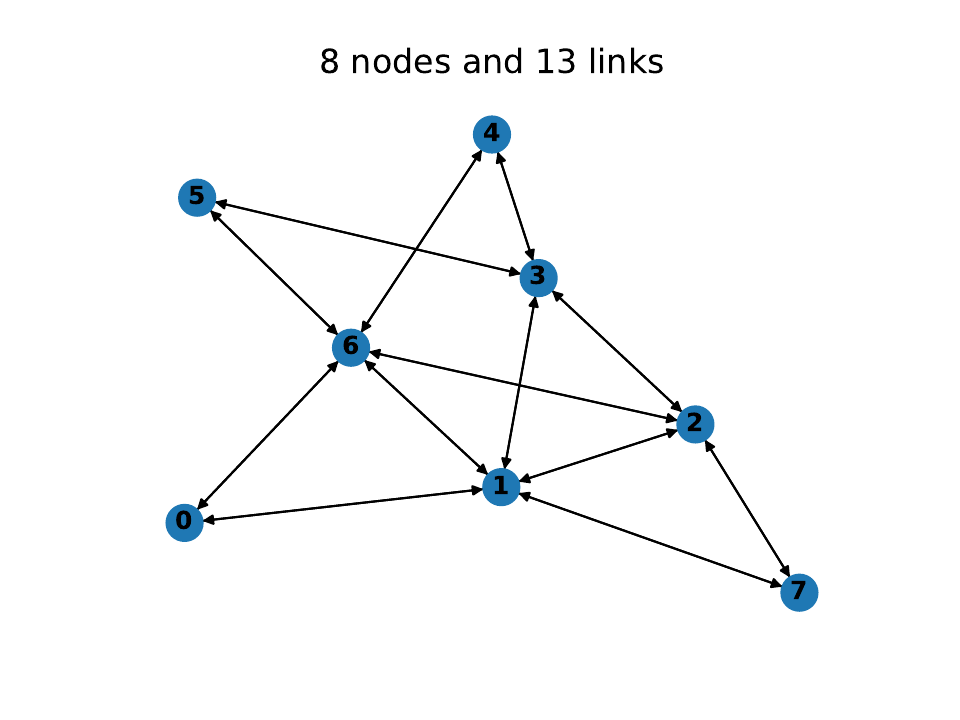}\hfill}
    \subfloat[]{\includegraphics[width=0.25\textwidth]{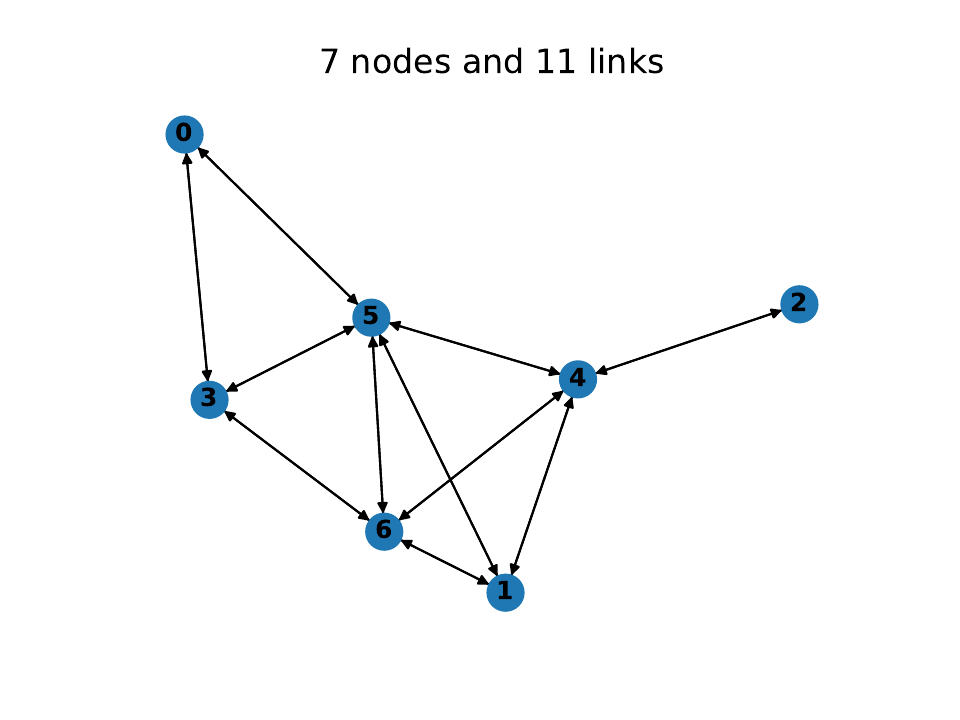}\hfill} \\
    \caption{Illustration of generated topologies.}
    \label{fig:topologies}
\end{figure*}

Figure~\ref{fig:topologies} shows the 11 topologies present in the network scenarios across the testbed.

\bibliographystyle{IEEEtran}
\bibliography{sample-base}

\newpage

\section*{Biography Section}
 




\begin{IEEEbiographynophoto}{Carlos Güemes-Palau}
(carlos.guemes@upc.edu) is a Ph.D. student at the Barcelona Neural Networking Center, Universitat
Politècnica de Catalunya.
\end{IEEEbiographynophoto}

\begin{IEEEbiographynophoto}{Miquel Ferriol-Galmés}
is a postdoctoral researcher at the Barcelona Neural Networking Center, Universitat Politècnica de Catalunya (UPC), Barcelona, Spain. He received a Ph.D. in Computer Architecture, where his research focused on applying Graph Neural Networks to optimize, analyze, and model computer networks. Currently, his work explores the application of AI in cancer research, leveraging machine learning to advance diagnostics and treatment in healthcare.
\end{IEEEbiographynophoto}

\begin{IEEEbiographynophoto}{Jordi Paillisse-Vilanova}
is a postdoctoral researcher at the Barcelona Neural Networking Center, Universitat Politècnica de Catalunya.
\end{IEEEbiographynophoto}

\begin{IEEEbiographynophoto}{Albert Lopez-Brescó}
is an informatics engineering graduate and head of engineering at the Barcelona Neural Networking Center, Universitat Politècnica de Catalunya.
\end{IEEEbiographynophoto}

\begin{IEEEbiographynophoto}{Pere Barlet-Ros}
is a professor at Universitat Politècnica de Catalunya and scientific director at the Barcelona Neural Networking
Center.
\end{IEEEbiographynophoto}

\begin{IEEEbiographynophoto}{Albert Cabellos-Aparicio}
is a professor at Universitat Politècnica de Catalunya and director of the Barcelona Neural Networking Center.
\end{IEEEbiographynophoto}

\vfill

\end{document}